\documentclass[3p,11pt,review,sort&compress]{elsarticle} \usepackage[hyphens]{url}
\usepackage{lineno}  \usepackage{graphicx}
\usepackage{booktabs} \makeatletter
\def\ps@pprintTitle{ \let\@oddhead\@empty
 \let\@evenhead\@empty
 \def\@oddfoot{\it \hfill\today} \let\@evenfoot\@oddfoot}
\makeatother

\usepackage[T1]{fontenc}
\usepackage{lmodern}
\usepackage{amssymb,amsmath}
\newcommand*\patchAmsMathEnvironmentForLineno[1]{	\expandafter\let\csname old#1\expandafter\endcsname\csname #1\endcsname
	\expandafter\let\csname oldend#1\expandafter\endcsname\csname end#1\endcsname
	\renewenvironment{#1}	{\linenomath\csname old#1\endcsname}{\csname oldend#1\endcsname\endlinenomath}}\newcommand*\patchBothAmsMathEnvironmentsForLineno[1]{	\patchAmsMathEnvironmentForLineno{#1}\patchAmsMathEnvironmentForLineno{#1*}}\AtBeginDocument{	\patchBothAmsMathEnvironmentsForLineno{equation}	\patchBothAmsMathEnvironmentsForLineno{align}	\patchBothAmsMathEnvironmentsForLineno{flalign}	\patchBothAmsMathEnvironmentsForLineno{alignat}	\patchBothAmsMathEnvironmentsForLineno{gather}	\patchBothAmsMathEnvironmentsForLineno{multline}}

\usepackage{ifxetex,ifluatex}
\IfFileExists{upquote.sty}{\usepackage{upquote}}{}
\ifnum 0\ifxetex 1\fi\ifluatex 1\fi=0   \usepackage[utf8]{inputenc}
\else   \usepackage{fontspec}
  \ifxetex
    \usepackage{xltxtra,xunicode}
  \fi
  \defaultfontfeatures{Mapping=tex-text,Scale=MatchLowercase}
  
\fi
\IfFileExists{microtype.sty}{\usepackage{microtype}}{}
\usepackage{longtable}
\usepackage{graphicx}
\makeatletter
\def\maxwidth{\ifdim\Gin@nat@width>\linewidth\linewidth
\else\Gin@nat@width\fi}
\makeatother
\let\Oldincludegraphics\includegraphics
\renewcommand{\includegraphics}[1]{\Oldincludegraphics[width=\maxwidth]{#1}}
\ifxetex
  \usepackage[setpagesize=false,               unicode=false,               xetex]{hyperref}
\else
  \usepackage[unicode=true]{hyperref}
\fi
\hypersetup{breaklinks=true,
            bookmarks=true,
            pdfauthor={Jaime Ashander},
            pdftitle={Predicting evolutionary rescue via evolving plasticity in stochastic environments}
            colorlinks=true,
            urlcolor=blue,
            linkcolor=magenta,
            pdfborder={0 0 0}}
\begin{document}
\begin{frontmatter}
	\title{Predicting evolutionary rescue via evolving plasticity in stochastic environments}
	    																			
	\author[esp,cpb]{Jaime Ashander\corref{cor1}}
	\cortext[cor1]{Corresponding author}
	\ead{jashander (at) ucdavis (dot) edu}
	\author[cefecnrs]{Luis-Miguel Chevin}
	\author[esp,cpb]{Marissa L. Baskett}

			\address[esp]{Department of Environmental Science \& Policy}
	\address[cpb]{Center for Population Biology
		UC Davis\\
		Davis, CA 95616\\
		USA
	}
	\address[cefecnrs]{Centre d'Ecologie Fonctionnelle \& Evolutive (CEFE)\\
	CNRS\\
	Montpellier CEDEX 5\\
France}

\begin{abstract}			Phenotypic plasticity and its evolution may help evolutionary
		rescue (ER) in a novel and stressful environment, especially if
		environmental novelty reveals cryptic genetic variation that
		enables evolution of increased plasticity.
		However, the environmental stochasticity ubiquitous in natural
		systems may alter these predictions because high plasticity may
		amplify phenotype-environment mismatches.
		Although previous studies have highlighted this potential
		detrimental effect of plasticity in stochastic environments,
		they have not investigated how it affects extinction risk in the
		context of ER and with evolving plasticity.
		We investigate this question here
		by integrating stochastic demography with quantitative genetic
		theory in a model with simultaneous change in the mean and
		predictability (temporal autocorrelation) of the environment.
		We develop an approximate prediction of long-term
		persistence under the new pattern of environmental fluctuations,
		and compare it with numerical simulations for short- and
		long-term extinction risk.
		We find that reduced predictability increases extinction risk
		and reduces persistence because it increases stochastic load
		during rescue.
		This understanding of how stochastic demography, phenotypic plasticity, and
		evolution interact when evolution acts on cryptic genetic
		variation revealed in a novel environment can
		inform expectations for invasions,
		extinctions, or the emergence of chemical resistance in pests.
\end{abstract}
\pagestyle{empty}
\begin{keyword}
	Evolutionary Rescue, Phenotypic Plasticity, Baldwin Effect,
	Environmental Predictability, Stochastic Demography, Environmental
	Stochasticity, Cryptic Genetic Variation
\end{keyword}

\end{frontmatter}

\clearpage{}\section{Introduction}

Abrupt environmental change beyond species' tolerance boundaries
occurs both naturally and due to human-driven global change
\citep{Palumbi2001}.  Change affecting an entire population (or one unable to
disperse)
leaves two possibilities for persistence: adapt or acclimate, that is, genetic
evolution or phenotypic plasticity
\citep{Davis2005}. Adaptive responses after a shift in the
environment can prevent extinction if there is sufficient additive genetic
variation \citep{Gomulkiewicz1995}.
Such evolutionary rescue
(ER) takes time, however, and a declining population
may go extinct before evolutionary response leads to positive growth and recovery
of population size \citep{Carlson2014}. Response via
phenotypic plasticity may be faster, while also permitting survival in novel
environments and time for further evolution.

Evolution and plasticity thus inevitably interact. On one hand, perfectly adaptive
plasticity prevents selection on fixed genetic
characters \citep{deJong2005} and more generally
may reduce the strength of selection on a trait in predictable environments.
On the other hand, partially adaptive
plasticity, simply by increasing survival in the new environment, results in
more time for selection, and thus evolution, before extinction
\citep{Crispo2007,Ghalambor2007}.
Furthermore, plasticity may itself evolve if it varies genetically (GxE
interaction, \citep[][]{Via1985}).  Plasticity, when quantified as the slope
of a linear reaction norm, can theoretically evolve to become
transiently higher in new environments \citep{Lande2009}. This greater
plasticity among surviving lineages requires
that the environmental shift causes increased additive genetic
variance ($V_A$) of the trait under stabilizing selection due to plasticity
(i.e., that stress reveals ``cryptic'' $V_A$, which occurs in some cases
\citep[reviewed by][]{Hoffmann1999,Charmantier2005,McGuigan2009} although the
opposite pattern is also frequently found).
Furthermore, empirical observations of heightened plasticity in lineages surviving anthropogenic
disturbances like climate change \citep{Willis2008,Merila2014} or transcontinental introductions
\citep{Davidson2011}
agree with suggestions from deterministic theory that plasticity facilitates ER
\citep{Chevin2010}.

Both the evolution of plasticity \citep{Moran1992,Gavrilets1993} and extinction
risk \citep{Lewontin1969,Lande2003} depend on environmental variation.
For plasticity, variation in the environment favours evolution of plasticity if
an environmental cue reliably correlates with the environment that imposes
selection
\citep{Moran1992,Gavrilets1993}.  More precisely, the optimal
level of developmental plasticity matches the correlation between the
environment of development and that of selection (i.e., environmental
predictability), with any mismatch reducing the expected long-term fitness
\citep{Gavrilets1993}.
For extinction risk, long-run population growth determines long-term persistence, and
declines with increasing variance in environmental fluctuations in the growth
rate \citep{Lewontin1969}. When such fluctuations are positively autocorrelated,
they increase extinction risk by allowing for many successive generations of
negative growth \citep{Turelli1977}.  In contrast, autocorrelation in phenotypic
selection might decrease extinction risk, because it allows closer evolutionary
tracking of an optimum phenotype, thus increasing the mean population growth
rate \citep[][]{Lande1996}.
Importantly, the pattern of environmental stochasticity
affects not only the mean, but the whole distribution of population sizes.
In the context of ER, this implies that many
populations may go extinct, even when the expected population does not
\citep{Lewontin1969,Burger1995}.

These separate influences of stochasticity on plasticity's
evolution and on extinction
suggest the potential for stochasticity to reduce, or possibly reverse, the
adaptive role of plasticity during ER.
For instance, if predictability
is low and plasticity is high, environmental variation in mean fitness will be
large
(because excess plasticity causes overshoots of the
optimum; \citep[][]{Chevin2013a}). Such excessive
plasticity can lead to extinction \citep{Reed2010}.
For example, if the environment undergoes an abrupt change in
predictability, which can happen if its temporal autocorrelation rapidly changes,
phenotypic plasticity might become transiently maladaptive, which would not only
reduce the expected fitness, but also increase the variance in population sizes
across replicates, further increasing extinction risk (as shown without
plasticity by Ashander and Chevin \textit{in prep}).
Therefore, considering environmental stochasticity is necessary to understand
the conditions under which evolving plasticity enhances or impedes ER.
Yet previous analytical treatments
\citep[e.g.,][]{Gomulkiewicz1995,Chevin2010} have neglected stochastic effects
on ER, which has rarely been studied outside of simulation models \cite[e.g.,][]{Bjorklund2009}.
Furthermore, for plasticity to evolve at all requires GxE interactions, which
with linear reaction norms may lead to
higher phenotypic variance in novel environments \citep{Gavrilets1993,Ghalambor2007,Lande2009}.
Large phenotypic variation in a new environment, for a trait
under stabilizing selection, results in standing variance load that reduces
population growth, which may prevent long-term persistence and thus impede ER
(see \citep[][]{Chevin2010}, Fig. 1c at large $t$) but whether these effects occur
in stochastic environments is unknown.

Here, we investigate whether and how stochastic environmental fluctuations,
and the variance load induced by expression of cryptic genetic variance in a novel environment, constrain evolutionary rescue with evolving plasticity.
To do so, we integrate quantitative genetic theory on evolution of plasticity
with stochastic demography.
Modelling a large shift in the mean optimum trait, to a value outside the
previous range of temporal environmental variation, combined with a change in
the environmental predictability of fluctuations in this optimum, we develop an
approximation for the population growth rate after the mean trait reaches a
stationary distribution around the expected optimum.
We also examine, using simulations of the underlying model, the risk of
quasi-extinction both in the short term and overall.
The approximations predicts long-term persistence in the new environment,
quantifying the eco-evolutionary dynamics that emerge with evolving plasticity
when a major detrimental environmental shift is combined with random
environmental fluctuations.
We find that for ER to occur in these conditions the environmental
predictability after the shift must be above a critical level.

\section{Materials and methods}

\subsection{Reaction norm, phenotypic selection, and population dynamics}\label{modelintro}

We assume random mating in a closed population
with discrete generations and environmental stochasticity
that is ``coarse-grained" so that every individual in a
generation experiences the same environment.
The environment both determines an optimal
value $\theta(t)$ for a primary trait $z(t)$, and cues a plastic response from that trait.
We assume linear dependence of the optimal trait on the
selecting environment $\varepsilon_s(t)$, so $\theta(t) = B \varepsilon_s(t)$ (where the
environmental sensitivity of selection $B$ defines the change in the optimum
phenotype for a unit change in environment $\varepsilon_s(t)$).
We model the genotypic reaction norm (i.e., plasticity), a linear response
in the trait to the environmental cue $\varepsilon_c(t)$ with slope $b$
and intercept $a$, such that the phenotype of an
individual is $z(t) = a + b \varepsilon_c(t) + e$.  Here the residual environmental
variation $e$ is independent of the macro environment and has mean zero and
variance $\sigma^2_e$ \citep{Lande2009}.
Our model applies to irreversible (non-labile) forms of plasticity such as
developmentally-plastic traits.

We model the reaction norm intercept $a$ and slope $b$ as quantitative
traits with means $\bar a$ and $\bar b$ and additive genetic variances $\sigma^2_a$ and  $\sigma^2_b$,
respectively \citep{Gavrilets1993,Lande2009}, so the intercept $a$
represents the breeding value in a reference environment,
$\varepsilon_c(t)=0$.  In addition, we assume that the population has
evolved in a range of environments centred around zero,
so that phenotypic variance in the reference environment is minimal.  Then with
linear reaction norms as here the slope
and intercept have
zero additive genetic covariance \citep{Lande2009}, and the additive genetic
variance of the expressed trait $z(t)$ increases quadratically away from the
reference environment
$\varepsilon_c(t) = 0$ (grey band in Figure~\ref{Fexplain}{\it (a)}),
which implies strong increases in heritability away from the reference
environment.
The mean and variance of the expressed trait value $z(t)$
before selection are
\begin{subequations}\label{Ez}
	\begin{align}
		&\bar z(t) = \bar a(t) + \bar b(t) \varepsilon_c(t)\\
  &\sigma_z^2(\varepsilon_c(t)) = \sigma^2_a + \sigma^2_b \varepsilon_c^2(t) +
  \sigma_e^2\label{Ezvar},
	\end{align}
\end{subequations}
which assumes the reference additive genetic variances are constant in time.
The expressed trait and the slope have covariance ${\rm Cov}(z,b) =
\varepsilon_c(t) \sigma_b^2$ and so,
with large $\varepsilon_c(t)$, direct selection on the trait results in stronger
selection on reaction norm slope \citep{Lande2009}.
(This does not hold with an alternative assumption where variance decreases away
from the reference environment; see~\ref{A:otherRN}.)

In each generation, we assume stabilizing selection for an optimum value,
which gives an expression for the Malthusian growth rate.
Given a fitness function of width $\omega$ and optimum $\theta$ defined above, absolute fitness is
$W(z,t) = W_{\max}\exp \left ( - \frac{(z(t) - \theta(t))^2}{2\omega^2}\right)$.
Averaging over the (normal)
distribution of phenotypes within a generation,
mean maladaptation of the trait from the optimum, $x(t) = \bar z(t) - \theta(t)$,
drives mean absolute fitness \citep{Lande1976a,Lande2009}
	$\bar{W}(t) = W_{\max}  \sqrt{S(\varepsilon_c(t)) \omega^2}
\exp \left ( - \frac{S(\varepsilon_c(t))}{2} \bar x (t) ^2
\right)$,
where $S(\varepsilon_c(t)) = \frac{1}{\sigma_z^2(\varepsilon_c(t))
\omega^2}$ is the strength of stabilizing selection experienced by
the population.
Because the strength of selection $S$ depends on the phenotypic variance,
it changes with the environmental shift according to eq.~\eqref{Ezvar};
however, in the new environment it is approximately constant (if selection is
weak as we assume here) with value
${S}_\delta$ as we show below
in \ref{Sapproxgr} (assuming small covariance between reaction norm intercept
and environment, see \citep[][]{Tufto2015}).
Trait change in a generation is the product of the additive genetic
variance-covariance matrix $\mathbf G$ and the selection gradient $\boldsymbol
\beta$, i.e., $\Delta \mathbf y(t) = \mathbf G \boldsymbol \beta$, with
the vector of reaction norm traits
$\mathbf y(t) = \left(\begin{smallmatrix}\bar a\\ \bar b\end{smallmatrix}\right)(t)$;
following Lande \citep{Lande1976a}, the gradient of log mean fitness $\bar W$ with
respect to $\mathbf y$ gives $\boldsymbol \beta$ (see~\ref{A:Ez}).
The dynamics for population size (assuming very weak density dependent
regulation, or a form of density-dependence, e.g., large exponent in a
theta-logistic model, where growth trajectories during rescue are similar; see
\citep[][]{Chevin2010})
are then $N(t+1) = \bar W(t)N(t)$, and the population's Malthusian growth rate
$r(t) = \log(\tfrac{N(t+1)}{N(t)}) = \log (\bar W(t))$ is
\begin{equation}\label{Er}
r(t) = r_{\max} + \log \sqrt{{S}_\delta \omega^2}
  - \frac{{S}_\delta}{2}
  x (t) ^ 2
  \end{equation}
Two terms reduce the growth rate from its maximum $r_{\max} = \log W_{\max}$.
They correspond to ``loads'' well-known in evolutionary biology. The first
is standing variance load due to phenotypic variability in any generation, the
second is the lag load \citep{MaynardSmith1976} due to maladaptation $x(t)$.
The latter can be further decomposed into a ``stochastic load'' caused by
fluctuations in the optimum, causing mismatch (with average zero) between the mean
trait and the optimum and a deterministic
``shift load'' caused by mismatch of the mean trait (after accounting for
fluctuations) relative to the mean optimum.

\subsection{Environmental stochasticity, shift in the optimum, and resulting
dynamics}

We consider an autocorrelated stochastic environment, where noise
arises from a stationary process $\varepsilon(t)$ with
variance $\sigma^2$ and autocorrelation function $f_\varepsilon(t; T) = e^{-t/T}$ (here,
$T$ is the characteristic autocorrelation time and $T \to 0$ implies the process
is white noise). The environment of development
determines the trait with a delay of $\tau$ less than one
generation, so $\varepsilon(t)$ determines both the environment of selection
$\varepsilon_s(t) = \varepsilon(t)$ and that of development
$\varepsilon_c(t) = \varepsilon(t - \tau)$, which is the environmental cue
\citep{Lande2009}.
(In general, environmental variables acting as cue might differ from the
environments causing selection, but this is beyond our present scope.)
In the reference environment, the correlation between the
cuing environment and the selecting environment represents the environmental
predictability of selection (or cue reliability;
\citep[][]{deJong1999,Reed2010}), which in our case is the
autocorrelation over time $\tau$, given by $\rho = f_\varepsilon(\tau) = e ^ {-\tau / T}$.

\begin{figure}[htpb]
	\begin{center}
		\includegraphics{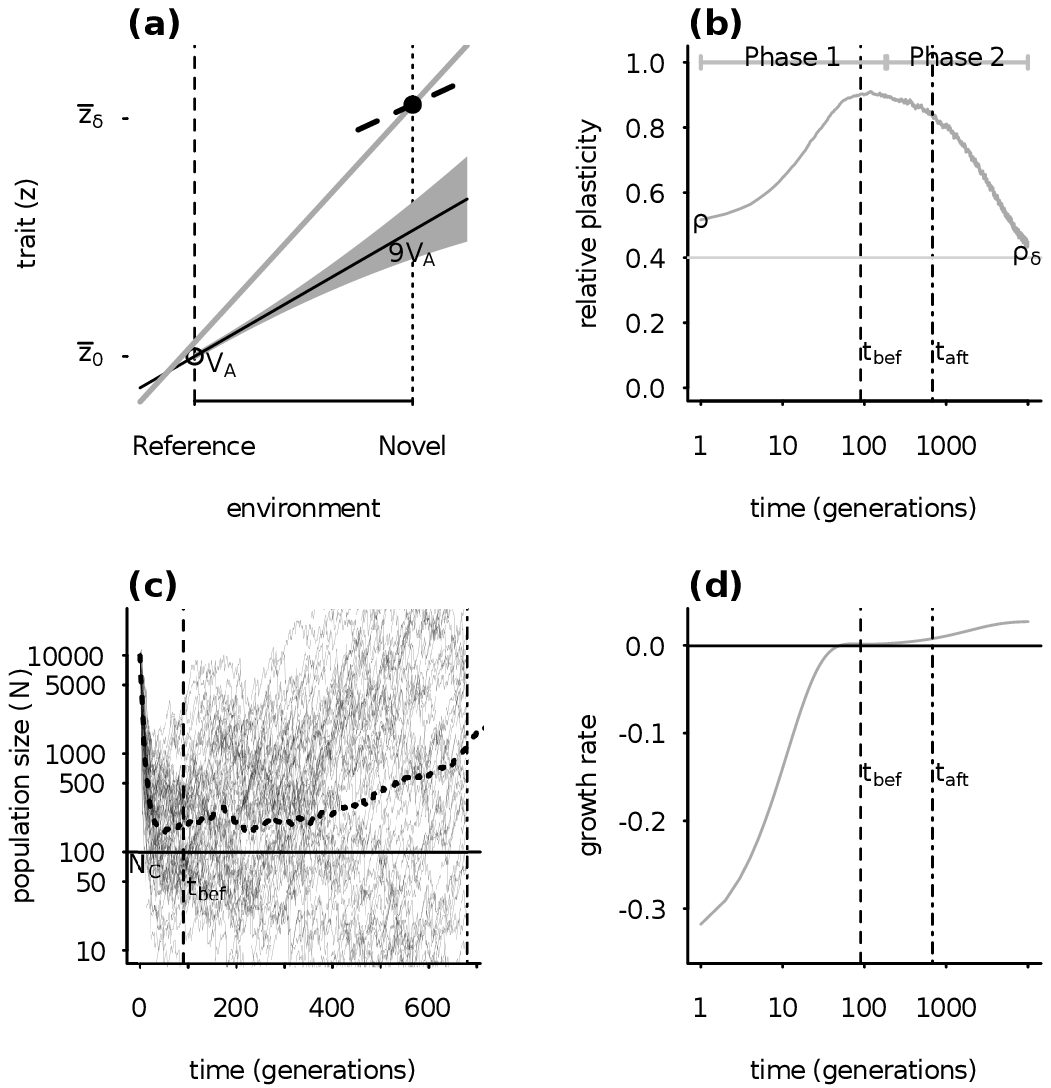}
	\caption[Two-phase adaptation and metrics]{Two-phase adaptation and
		metrics for extinction avoidance and persistence. {\it (a)}
		Our scenario: shifting the mean environment (dashed to dotted vertical
		line) alters the mean optimum trait (from open circle to solid
		dot), while the change in environment autocorrelation changes
		the optimal plasticity (from slope
		of solid black line to that of dashed black line). Due to GxE,
		additive variance increases in the new environment (by a factor
		of approximately 9 for these
		parameters, grey band; see {\it Discussion} for empirical
		examples where such inflation may occur). During
		evolutionary rescue, the mean reaction norm (solid back line with slope
		initially evolved to match predictability $\rho$) increases
		transiently (grey line; note the intercept increases a small amount also)
		and eventually evolves to match the
		new predictability $\rho_\delta$ (slope of dashed black line). {\it (b)}
		Change occurs in two phases.  Mean relative plasticity ($\bar b
		/ B$; versus time, log scale) increases quickly during Phase
		1, then decays slowly
		during Phase 2.  Over the transition from Phase 1 to
		Phase 2 (from $t_{\rm bef}$, dash vertical line, to $t_{\rm
			aft}$, dash-dot vertical line), plasticity is relatively constant.
		{\it (c)} Population size versus time in simulations, mean size
		(grey line) declines during Phase 1 of rescue, while variance
		(thin lines) increases, heightening extinction risk.  We compute the
		probability of quasi-extinction before rescue at
		$t_{\rm bef}$ and at $t_{\rm aft}$ as the
		proportion of simulated trajectories below $N_C =
		$100 (horizontal black bar).  {\it (d)} The
		mean population growth rate (versus time, log scale) is initially
		negative, increases during Phase 1 and is relatively stable
		during Phase 2 (grey line).
		We define persistence as
		self-replacement after Phase 1, using growth rate predicted at
		the Phase's end
		(solid horizontal line, $\bar r_1 \ge 0$, eqn \ref{Eltpersist};
		note when accounting for both Phases, the growth rate slowly
		increase during Phase 2: grey line).  Parameters: shift size
		$\delta=$
		4, predictability before $\rho =$ 0.5
		and after the shift $\rho_\delta=$ 0.4, selection
		strength $\omega^2=$ 20, developmental delay
		$\tau=$ 0.2, additive genetic and environmental
		variances $\sigma_a^2=$ 0.1, $\sigma_b^2=$
		0.05, and $\sigma_e^2=$ 0.5; initial
		population size $N(0)=$ \ensuremath{10^{4}}, and maximum fitness
		$e^{r_{\max}}=$ 1.1.
	\label{Fexplain}}
\end{center}
\end{figure}

A single discrete shift in the environment changes the mean
of both the environment of selection and the environmental cue by the same
amount $\delta$,
and also changes the environmental predictability from $\rho$ to $\rho_\delta$.
Our analysis assumes that, even accounting for stochastic variance in the environment, the new
optimum is very different from the previous one, i.e., $\sigma^2 \ll \delta^2$.
Under this assumption, rescue occurs over two phases at different
timescales \citep{Lande2009}. Over the rapid Phase 1, the maladaptation of the
mean phenotype is reduced to near zero by evolution of increased plasticity
(mean reaction norm slope increases from solid black line to grey line in Figure
\ref{Fexplain}{\it (a)}, rapidly as in Figure~\ref{Fexplain}{\it (b)} left of the dotted lines).
During the slower Phase 2, the mean reaction norm height and slope evolve to
approximately their optimal values (mean reaction
norm slope decreases from solid grey line to dashed black line in Figure
\ref{Fexplain}{\it (a)}, slowly as in Figure~\ref{Fexplain}{\it (b)} right of the dotted lines).
Lande \citep{Lande2009} showed that the rapid increase of plasticity is controlled by
the proportion of additive genetic variance in the new environment due to
variance in reaction norm slopes,
$\phi = \frac{\delta ^2 \sigma _b^2}{\sigma _a^2+\delta ^2 \sigma _b^2}$, where
assuming a large shift implies $\phi$ is near $1$.
In the new environment, the maladapted population declines at first
(Figure~\ref{Fexplain}{\it (c)}) because it has a negative expected growth
rate, which increases over time due to adaptive evolution (eq.~\ref{Er}; Figure
\ref{Fexplain}{\it (d)}).
Because Phase 1 occurs much faster than Phase 2, the growth rate first
increases rapidly during this phase, then effectively stabilizes.

\subsection{Simulations and analysis}

We quantify the risk of extinction before rescue (i.e., before the end of Phase 1) by using simulations to
compute the proportion of trajectories that experience quasi-extinction.
Additionally, we quantify two components of extinction risk, which reflect the
action of differing eco-evolutionary processes.
First, the short-term risk reflects both temporally stochastic environment and
deterministic effects of reduced shift load during ER.
We compute it by using simulated quasi-extinction  during the initial population
decline before the end of Phase 1.
Second, the long-run growth rate reflects effects of stochastic load and standing
variance load at stationarity; a negative long-run growth rate implies eventual extinction.
We analytically approximate the long-run growth rate at the end of Phase 1, and also
we compute it from simulations for comparison.

\subsubsection{Simulations of short-term and total extinction risk}

For all simulations of extinction risk, we drew initial
conditions from a stationary distribution of reaction norm intercept
and slope generated after 15,000 generations in an environment with mean $0$ and
predictability $\rho$ (where theory predicts $\bar a(t) = 0$ and slope $\bar
b(t) /B = \rho$; see \citep[][]{Gavrilets1993,Lande2009}).
Using mean fitness as growth rate, we compute population dynamics as $N(t+1) = \bar W(t)N(t)$.
We also track the reaction norm parameters $\bar a, \bar b$, and mean fitness $\bar W$, with trait
change in a generation given by the product of genetic variance (assumed
constant, at an equilibrium between mutation and stabilizing selection) and the
selection gradient (see~\ref{modelintro}).  We computed
means of these quantities at each generation across 250 replicate simulations
(except in Figure~\ref{Fexplain}, where we used 50 replicates to ease
visualization).
We also performed simulations where we relaxed the assumption of constant
variance, assuming instead that variance reaches an equilibrium at each
population size due to mutation-selection-drift balance,
(under a modified stochastic house-of-cards approximation
\citep[][]{Burger1995}; see Figure~\ref{FS6potrescuesmalldelta}).
All numerical simulations and plotting of analytical predictions were performed
in R \citep{RCoreTeam2014,Eddelbuettel2011,Ashander2016}; further details
are in~\ref{A:envsim}.

We define quasi-extinction probability $P_{\rm QE}(t)$ as the
proportion of population trajectories that fall below a critical population size
$N_C$ before time $t$. From the simulations, we computed quasi-extinction
probability for the short-term, at a time before the end of Phase 1
($P_{\rm QE}(t_{\rm bef})$, with $t_{\rm bef}$ as half the characteristic timescale of Phase
2).
Also from the simulations, we computed the total probability of quasi-extinction,
over the whole time period into Phase 2 ($P_{\rm QE}(t_{\rm aft})$, see
Figure~\ref{Fexplain}{\it (c)}; we defined $t_{\rm aft}$ as the characteristic timescale
of Phase 2 plus 500 generations).

\subsubsection{Analysis of the approximate long-run growth rate}\label{Sapproxgr}

A positive long-run growth rate (asymptote of the black line in Figure~\ref{Fexplain}{\it (d)})
is necessary for long-term persistence after rescue.
The long-run growth rate $\bar r(t)$ depends on the expectations of the standing variance and lag loads,
taken over stochastic fluctuations, of log mean absolute fitness \citep{Lewontin1969,Burger1995,Lande1996}.
An analytical formula for growth rate follows from two main assumptions, under
which we compute the expected loads (derived in~\ref{A:approxdyn}).
First, if stabilizing selection is weak,
by taking an expectation over stochastic fluctuations the variance load caused by increased phenotypic variance is approximately
constant, with value $L_G = - 1/2 \log ({S}_\delta \omega^2)$,
where ${S}_\delta\approx \left ( \sigma_a^2 + \sigma_b^2 (\delta^2 + \sigma^2) + \sigma_e^2 + \omega^2 \right)^{-1}$,
is the average selection strength in the new environment.
Also assuming weak stabilizing selection, the expected lag load is approximately
$L_L(t) = \frac{{S}_\delta}{2} \mathbb E(x ^2 (t))$
and is the only quantity that varies in time.
Thus, the lag load determines temporal variation
in the long-run growth rate, which is $\bar r(t) \approx r_{\max} - L_G - L_L(t)$.

A second assumption is that fluctuations in maladaptation achieve stationarity
at the end of Phase 1, which permits us to compute the variance in maladaptation
$\sigma_x^2$.
Because the lag load can be decomposed into the mean and
variance in maladaptation, $L_L(t) = \frac{{S}_\delta}{2} (\bar x ^2 (t) +
\sigma_x^2$ and the mean maladaptation goes to zero by the end of Phase 1
(see~\ref{A:Ez}), the long-run growth rate depends only on the variance in maladaptation
$\sigma_x^2$, and is
$\bar r(t) \approx r_{\max} - L_G - \frac{{S}_\delta}{2} \sigma_x^2$.
The variance in maladaptation affects not only the spread in trajectories of growth rates,
and thus population size, but also long-term persistence, based on the long-run growth rate. 
After Phase 1, change in reaction norm parameters is slow which provides some
justification for the assumption, which yields an explicit formula for the variance
in maladaptation, and thus the growth rate. The formula depends on 
the value of plasticity is at the end of Phase 1 (with value $b_{\max} = B(\rho
+\phi (1 -\rho))$), as well as characteristics of the novel environment (see
full derivation in~\ref{A:ELs}).
Persistence occurs when
\begin{equation}\label{Eltpersist}
	\bar r_1 = r_{\max} - L_G -
	\frac{{S}_\delta}{2} \left (
	\frac{
	\sigma^2 ( B^2  + \bar b_{\max} (\bar b_{\max} - 2 B \rho_\delta))
	}
	{1 - {S}_\delta  \sigma_a^2
	\log \left [ \rho_\delta^{1/\tau}
	\left ( 1 - \frac{B \bar b_{\max} (\rho_\delta^{-1} - \rho_\delta)}
		{B ^2 + \bar b_{\max} (\bar b_{\max} - 2 B \rho_\delta)} \right) \right ]^{-1}
	}
\right)
	\ge 0.
\end{equation}
Because $\bar r_1$ depends on environmental predictability $\rho_\delta$ in the new
environment, the size of shift in mean $\delta$, and (through selection strength
${S}_\delta$) variance in plasticity $\sigma_b^2$, eq.~\eqref{Eltpersist}
defines critical levels of these that are necessary for a population to persist.
For comparison to the analytically-predicted critical parameter values
 ($\bar r_1 \ge 0$ eq.~\ref{Eltpersist}), we also computed (from the
 simulations described above) the stochastic population growth rate
over the cusp between Phases 1 and 2 (i.e., between $t_{\rm bef}$ to
$t_{\rm aft}$ Figure~\ref{Fexplain}{\it (d)})
from its maximum likelihood estimator,
$\hat \lambda_s = \frac{N(t_{\rm bef}) - N(t_{\rm aft})}{t_{\rm bef}- t_{\rm aft}}$
\citep[][]{Caswell2001}.

\section{Results}

We found that with evolving plasticity, the combination of a major environmental
shift with stochastic fluctuations in the environment alters the eco-evolutionary
outcome, as compared to the effects of each of these factors in isolation.
In particular, our analysis reveals the importance of environmental
predictability for ER with evolving plasticity.

\subsection{Environmental predictability is critical to ER with evolving plasticity}

Evolutionary rescue following a shift in the mean optimal trait requires a critical
level of final environmental predictability.
Predictability below
this critical level both reduces long-term persistence
(Figure~\ref{Fpotrescue}{\it (a-b)} and
increases extinction risk in the short term (Figure~\ref{Fpotrescue}{\it (c,d)} .
Furthermore, this decreased predictability strongly increases total extinction risk across a range
of initial genetic variances in plasticity (Figure~\ref{Fpotrescue}{\it (e)})
and sizes of the environmental shift (Figure~\ref{Fpotrescue}{\it (f)}).
None of these effects of predictability can be understood from deterministic
models, which predict inaccurate trajectories over rescue in the presence of
stochasticity (see Figure~\ref{FS4dynamics}).

Increases in stochastic load, due to increased plasticity during ER, cause this
constraint by reducing the long-run growth rate.
For any fixed shift size
($\delta$ Figure~\ref{Fpotrescue}{\it (b)}) or additive variance in plasticity
($\sigma_b^2$ Figure~\ref{Fpotrescue}{\it (a)}), decreasing the final environmental predictability
$\rho_\delta$ increases the stochastic load because plasticity in excess of
predictability causes the mean trait to overshoot the mean optimum (see~\ref{A:ELs} and Figure~\ref{FS2stochload})
resulting in larger mismatch variance and hence
decreased expected growth rate after Phase 1.
It is important to note that a reduction in predictability (i.e., temporal autocorrelation in
the optimum) is expected to increase stochastic load even without plasticity,
because it will decrease adaptive tracking \citep[][]{Lande1996}.
In our case, however, evolved increases in plasticity causes much higher
stochastic load (often more than four times greater, Figure~\ref{FS2stochload}).
In simulations where the genetic variance changes with population size due to
drift, there is still a
critical predictability (see Figure~\ref{FS7potrescuevg}) but it increases much
faster with shift size $\delta$ and the effect of standing variance load
disappears.

\subsection{Effects of genetic variance in plasticity depend on initial
	plasticity}

\begin{figure}[htbp]
	\begin{center}
		\includegraphics{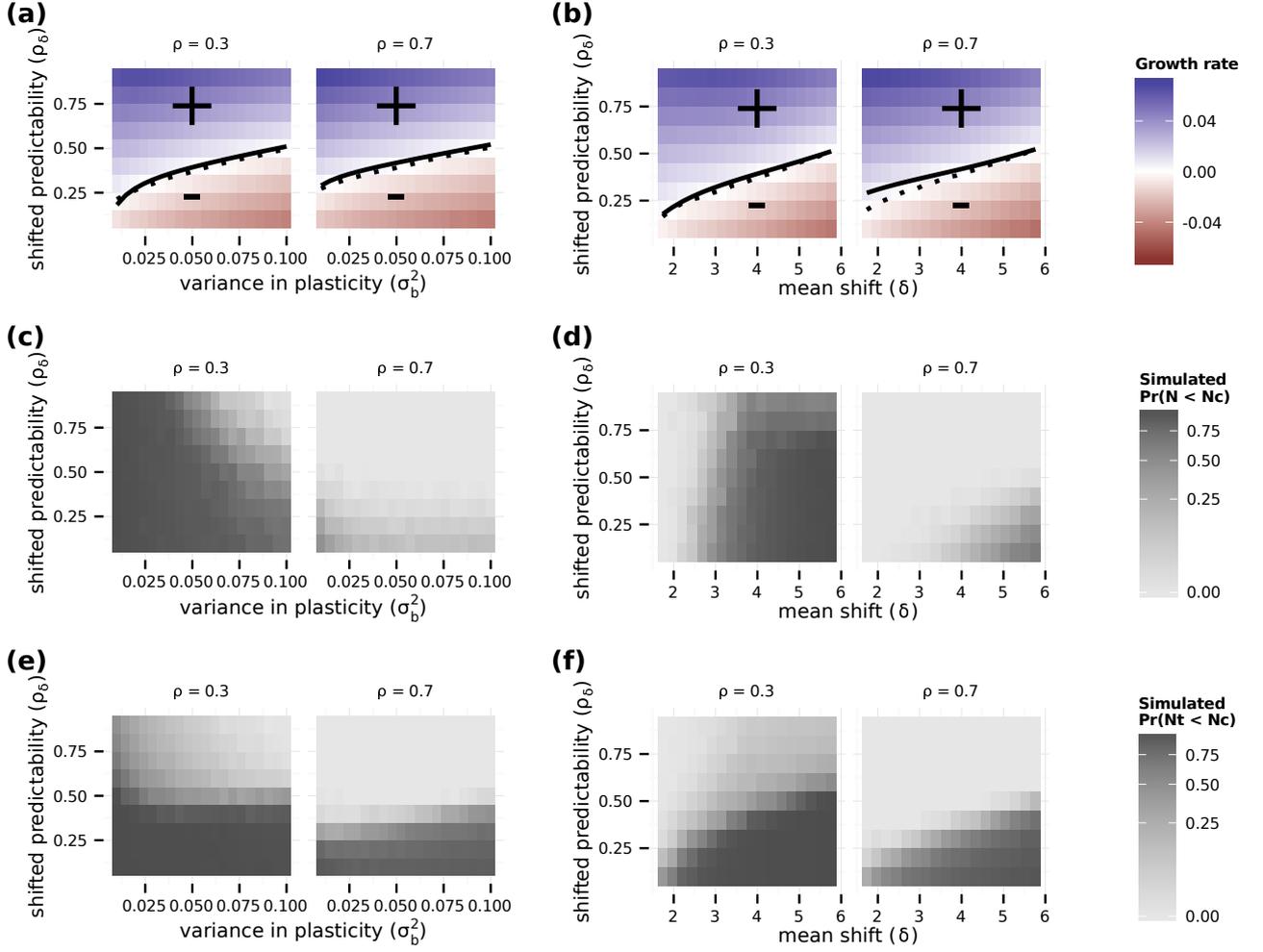}
	\caption[Predictability, additive variance in plasticity, and mean shift]{
		Potential for evolutionary rescue over a range of values for
		post-shift predictability $\rho_\delta$ versus genetic variance
		in plasticity the reference environment $\sigma_b^2$ {\it (a,c,e)}
		and shift size $\delta$ {\it (b,d,f)}. Within panels, columns show
		low ($\rho =$ 0.3) and high  ($\rho =$
		0.7) initial predictability.  {\it (a,b)} Growth
		rates at the end of rescue are computed from numerical
		simulations as the stochastic growth rate $\lambda_s$ between
		$t_{\rm bef}$ and $t_{\rm aft}$ spanning Phase 1 and 2
		(diverging heatmap: white 0, blue positive, and red negative).
		Black lines indicate the threshold between decline (-) and
		persistence (+) based on the analytical approximation
		($\bar r_1 = 0$, eqn~\ref{Eltpersist} ; solid line) and
		stochastic simulations ($\lambda_s = 0$, dotted black line).
		{\it (c-f)} Simulated probability of quasi-extinction
		at a point during rescue (i.e, at $t_{\rm bef}$, {\it (c-d)} and total
		(i.e., up to $t_{\rm aft}$, {\it (e,f)} darker colors indicate
		greater chance of extinction (grayscale heatmap).
		Quasi-extinction is the proportion of trajectories below a
		critical size $N_C$ by a given time (both at $t_{\rm bef}$
		or $t_{\rm aft}$; illustrated in Figure~\ref{Fexplain}).
						When variance in plasticity is varied {\it (a,c,e)}, shift size is
		set to $\delta=$ 4 (so that additive
		variance increases by a factor of 9 in
		the new environment when $\sigma_b^2 =$0.05).  When shift size is
		varied {\it (b,d,f)}, variance is set to $\sigma_b^2=$
		0.05.  Other parameters: initial population size
		$N(0)=$ \ensuremath{10^{4}}, selection strength $\omega^2=$
		20, developmental delay $\tau=$ 0.2,
		additive genetic $\sigma_a^2=$ 0.1 and environmental
		$\sigma_e^2=$ 0.5 variances, maximum fitness
		$e^{r_{\max}}=$ 1.1, and variance in the
		environment $\sigma^2 = 1$.
		\label{Fpotrescue}}
\end{center}
\end{figure}

In populations with low initial plasticity, determined by the environmental
predictability $\rho$ with which a lineage has evolved,
increasing genetic variance in plasticity can greatly reduce short-term risk of
quasi-extinction (for $\rho = 0.3$ Figure~\ref{Fpotrescue}{\it (c,d)}
left column, $P_{\rm QE}(t_{\rm bef}) > 0.75$ for all $\sigma_b^2$ below about
0.025).
Effects on the total risk of extinction are similar (Figure~\ref{Fpotrescue}{\it
	(c)}, left column).

On the other hand, in populations with high initial plasticity, genetic variance
in reaction norm slope
does not affect extinction very much in the short term ($P_{\rm QE}(t_{\rm bef}) < 0.05$
for $\sigma_b^2$ below 0.025 with high predictability $\rho_\delta$ in
Figure~\ref{Fpotrescue}{\it (c,e)} right column).
Such populations generally have much lower risk of
short-term quasi-extinction consistent with earlier results of Chevin and Lande \citep[][their
Fig. 2]{Chevin2010} on the effect of initial relative plasticity on
(deterministic) extinction.
However, in these populations when predictability following the shift is intermediate to low, increasing additive
variance in reaction norm slopes decreases long-run growth rate (positively
sloped solid lines in Figure~\ref{Fpotrescue}{\it (a)} indicate increased variance moves
from `+' to `-', with similar increase in the total risk extinction
(Figure~\ref{Fpotrescue}{\it (c)}, right column).

In low-plasticity populations transitioning to
high predictability environments, the effects of stochastic load are low and the
benefits from increased plasticity during ER can outweigh the negative effects
of the standing variance load.
In contrast, for populations with initially high plasticity the change in
plasticity during ER is small and the effects of standing variance load
and stochastic load dominate.
This can be seen by comparing the analytical persistence threshold
(Figure~\ref{Fpotrescue}{\it (a-b)}) to the total extinction probability
(Figure~\ref{Fpotrescue}{\it (e,f)}).
For high initial plasticity (right columns) the analytical prediction matches
the total extinction probability, but this is not the case for low initial
plasticity (left columns).
The equation predicts the same constraint
in both low- and high-plasticity populations (solid lines have similar
shape in both columns of Figure~\ref{Fpotrescue}{\it (a-b)}).
Numerical simulations of long-run growth rate agree (heatmap and dotted line have
same shape in both columns of Figure~\ref{Fpotrescue}{\it (a-b)}).
This occurs because eq.~\eqref{Eltpersist} depends on plasticity at
the end of Phase 1, which is influenced very little by initial plasticity.
The condition of positive long-term growth, however, is only necessary for ER,
not sufficient, and is sensitive only to effects of variance load and stochastic
load. Total extinction risk reflects deterministic reduction of lag load due to
increase plasticity during ER; such increases are more important for
low-plasticity populations, which incur greater maladaptation initially.

\subsection{Analytical prediction of the growth rate performs well}

Across most parameter values agreement between eq.~\eqref{Eltpersist} and
simulations is strong (compare solid and dotted lines within Figures
\ref{Fpotrescue}{\it (a-b)}).
The exception is high initial plasticity and small environmental shift
(solid and dotted lines mismatch for low predictability in Figure
\ref{Fpotrescue}{\it (b)}), where
changes in $\phi$ are driven by small shifts $\delta$ and large additive
variance in plasticity in the reference environment (see~\ref{A:approxbad}).

\section{Discussion}

Our analytical results and simulations reveal the eco-evolutionary dynamics that emerge with evolving plasticity when a major detrimental environmental shift is combined with random environmental fluctuations.
We find that whether evolving plasticity will enable ER depends on environmental predictability in the new environment.
If predictability is moderate to low after an environmental shift, the transient
evolution of high plasticity that occurs in the new environment causes a large
stochastic load that reduces the likelihood of ER.
Even without plasticity, environmental predictability affects the stochastic load
(lower autocorrelation reduces adaptive tracking of the optimum by genetic
evolution; \citep[][]{Lande1996}), and thus the probability of evolutionary rescue in a
stressful new environment (Ashander and Chevin \textit{in prep}).
When ER causes increased
plasticity, these effects are stronger (see Figure~\ref{FS2stochload}), because frequent
mismatches caused by excess plasticity result in large variance in growth rate
and negative population growth (Figure~\ref{FS4dynamics}{\it (d)}), even after mean
maladaptation has reduced to zero (Figure~\ref{FS4dynamics}{\it (a)}).
This parallels findings that non-evolving plasticity can amplify fluctuations in population mean fitness
and thus growth rate without an environmental shift \citep{Reed2010,Chevin2013a,Michel2014};
here, we demonstrate that this process also constrains ER.
On the other hand, if predictability is high in the new environment, ER is relatively likely.
Our findings accord with other theory that suggests irreversible developmental plasticity is not
useful in low predictability environments, which might instead favour reversible
plasticity \citep{Botero2015}.
In addition, we find that positive effects of increased genetic variance in
plasticity for ER in a stochastic environments are limited to situations where
lineages with low evolved plasticity experiencing shifts to a more predictable
environment.

Due to the trade-off between short-term adaptive benefits and
long-term stochastic and standing variance loads, large genetic variance in plasticity
can sometimes decrease the chance of ER for lineages where plasticity is
initially high that experience shifts to environments with low predictability.
Chevin and Lande \citep[][their Fig. 1c at large $t$]{Chevin2010} noted the
effect of the standing variance load, but
focused on deterministic environments that do not include random noise.
We extend this theory to noisy environments and demonstrate, for low
predictability, the effects of genetic variance in
plasticity: it lessens exposure to small population size in the short term, but
may cost increased variance load that slightly reduces long-term persistence (in
population with already-high plasticity that shift to low-predictability
environments).
(Note, however, that in our model the genetic variance in plasticity is fixed,
and so the model cannot produce any direct selection to reduce this load.)
These effects of genetic variance, however, are weak compared to
the constraint imposed by predictability.

Our findings extend deterministic theories on ER
\citep[e.g.,][]{Gomulkiewicz1995,Chevin2010} by directly quantifying the effect
of environmental stochasticity with evolving plasticity on persistence.
Although models of ER on quantitative traits have not typically accounted for
environmental stochasticity \citep[but see][]{Burger1995,Bjorklund2009}, demographic
stochasticity has been shown to affect ER by \textit{de novo} mutation or
standing variation at a single gene
\citep[e.g.,][]{Martin2013a} because population trajectories
depends on births and deaths during the initial period when advantageous genes
were rare.
Environmental stochasticity, our focus here, is arguably more
important than demographic stochasticity because it operates with equal strength at all
population sizes \citep{Lande2003}, and affects the population size distribution
during ER or with fluctuating selection on quantitative traits (Ashander and Chevin \textit{in prep}).
Stochasticity's effect on ER with evolving plasticity may be especially strong,
as mismatches of increased plasticity with predictability both increase
short-term extinction risk and reduce long-run growth.
We also showed that lowered predictability can cause the high plasticity that evolved
during ER to eventually be maladaptive, unless the new environment is highly
predictable.

\subsection{Assumptions and caveats}

To obtain analytical approximations that predict evolutionary trajectories
we used three main assumptions.  We assumed first,
that baseline additive variances in reaction norm parameters remain constant
during
the shift, second, that the linear shape of reaction norms extend beyond the
reference environment where they evolved and where variance is minimized, and
third, that the new environment is far outside the distribution of past
environments and causes a density-independent decline in population size.

Constant additive genetic variance, as modelled in our simulations and
analytical results, is commonly assumed in models like
ours to make analytical progress, and although it is not biologically realistic
it can provide a good approximation to more complex dynamics
\citep{Turelli1994}.
Accounting for evolving genetic variance would require added complexity, such as
tracking the full distribution of breeding values:
\citep[e.g.,][]{Burgess2013} or using an approximation like the stochastic
house-of-cards \citep[e.g.,][]{Burger1995}.
More explicit genetics have already been included in some models of ER
(e.g., polygenic adaptation, \citep[][]{Gomulkiewicz2010}), but this is more
challenging with plasticity and a stochastic environment.
With environmental noise in a constant environment, variance in reaction norms is expected to decrease
\citep{deJong2000}
which could represent the state of the population in the long run after the
phenotype has evolved to become canalized around the new mean environment
\citep{Kawecki2000}
but theory is lacking for the transient change in genetic variance after the
shift in mean environment.
It is likely that reductions in population size during ER will reduce
genetic variance.
We investigated the sensitivity of our
results to this possibility (see Figure~\ref{FS7potrescuevg}) and still found
that if predictability in the new environment is below a critical level then ER
is unlikely.
The critical levels in this case, however are higher than those suggested by our
analytical results, which thus can be viewed as a lower bound on extinction risk.

The dynamics we show are
all derived by assuming that partially adaptive plasticity with linear reaction norms
extending
to the novel environment and that variance is minimal in the reference environment.
Theory demonstrates that
linear reaction norms will evolve within the reference regime over long
timescales \citep[e.g.,][]{Gavrilets1993}, but it is the assumption that
variance is minimal in the reference environment (and that reaction norms extend
to the novel environment, \citep[][]{Lande2009}) that implies increased heritable variation in the novel
environment.
Without increased additive genetic variance $V_A$, there is no strong
covariance between reaction norm slope and the trait expressed, which means no strong selection to increase reaction norm slope and
so no transient increase in plasticity (see~\ref{A:otherRN}). An environmental shift then
increases neither variance load nor, necessarily, plasticity.
Because we expect quite different results without assuming $V_A$ increases in
the new environment, we emphasize that our models will only apply when novel or stressful environments
reveal cryptic genetic variation \citep[e.g.,][]{Ghalambor2007}.
Although this idea finds support in some systems, in meta-analyses the opposite
trend (of decreasing heritable variation in
rare, stressful or novel environments) is equally frequent \citep[reviews:
][]{Hoffmann1999,Charmantier2005}.  There are relatively few studies that
obtain clear results either way, however, in part due to the difficulty of
replicating an experiment across many environmental values \citep{McGuigan2009}.
In these cases, then, evolution of plasticity should have little influence on ER
and we expect dynamics to follow results for non-plastic ER [e.g., classic
deterministic theory \citealp{Gomulkiewicz1995}, or its extension to
stochastic environments by Ashander and Chevin \textit{in prep}].
Furthermore, although we assumed partially adaptive plasticity, it can be sometimes be maladaptive
\citep[e.g.,][]{Duputie2015}.
Developing theory on this may require modelling non-linear reaction norms
(e.g., via function-valued traits, \citep[][]{Gomulkiewicz1992}).

Finally we ignore demographic regulation, effectively assuming that density-dependence is very
weak.
In practice we assume the novel environment is stressful and initially
causes a density-independent population decline because the environment is far outside the previous range of
environments.
Introduced by Lande \citep{Lande2009}, it is an extreme version of environmental
novelty, but one that yields mathematically tractable expressions for the growth
rate.
As shown previously without stochasticity, the density-independent trajectories
we study are close to those under very weak density dependent regulation
or a form of density-dependence, e.g., large exponent in a theta-logistic model,
where growth trajectories during rescue are similar \citep{Chevin2010}.
Under stronger density regulation that acts even when the population is far
below carrying capacity, we would expect steeper declines in population size.

Despite the limitations mentioned above, our assumptions apply nicely to some systems. For example,
compare the implied increase in $V_A$ ($\approx 9$-fold; Figure~\ref{Fexplain}{\it (a)}) to
Husby \textit{et al.} \citep{Husby2011} who showed higher temperature increased genetic variance
of breeding time of the great tit {\em Parus major} ($\approx
4$-fold increase in mean $V_A$), or to McGuigan \textit{et al.} \citep{McGuigan2011} who showed for
three-spine stickleback ({\it Gasterosteus aculeatus})
an even stronger increase in genetic variance of body size with
low-salinity ($\approx 38$-fold increase in mean $V_A$).
How frequently such increases in $V_A$ occur with environmental novelty
remains an open question. Furthermore, even if increases occur, they are
not sufficient to guarantee rescue. Among other conditions, heritability and
evolvability must also increase, and as we show heritable variance in plasticity
may be detrimental for several reasons (standing load, and stochastic lag load
in unpredictable environments).

\subsection{Empirical context and applications}

We define a long-term persistence criterion by predicting the stochastic growth rate for
hundreds to thousands of generations after the population's mean trait has
adapted to the new optimum.
To apply this theory, either for empirical verification or for prediction,
several types of data are needed.
First, estimates for parameters governing the genetics of GxE interactions (i.e., additive
genetic variances in several environments) and the trait's effect on
fitness can be obtained from common garden and other experiments
\citep{Merila2014}.
Second, information about the environmental sensitivity of selection can be
measured using a single episode of selection (e.g., in  {\em Parus
major}: \citep{Vedder2013,Gienapp2013,Chevin2015}). Third, parameters
for environmental predictability can be characterized statistically
\citep[e.g.,][]{Marshall2015}, but this requires knowledge, or assumptions,
about the environment of selection.

These data requirements are challenging but achievable in several field and
laboratory systems.  Phenological traits, because they are developmental traits
under strong selection and cues are often known, may be the best fit.
Furthermore, these traits among the most observable biological responses to
climate change \citep{Parmesan2006}.
Germination timing of high altitude plants is particularly
promising: winter temperature and snow melt cue development that is also
genetically-influenced and under stabilizing selection
(with risk of frost-killing if too early, or dessication if too
late; \citep[][]{Inouye2013}), and optimal timing varies with altitude
\citep{Anderson2012a},
so reciprocal transplants shift the optimum.
Furthermore, optimal timing is shifting with climate change
\citep{Bradshaw2008}.
Osmoregulatory traits are another candidate.
Studies in the copepod {\it Eurytemora affinis} support a role for rapid
evolution driven by plasticity in parallel adaptations to freshwater
\citep{Lee2011}. For three-spine stickleback
additive genetic variation in body size is higher in stressful low-salinity
environments \citep[][]{McGuigan2011}.
ER has already been studied in several microorganisms (e.g., {\it Pseudomona
	fluorescens}, \citep[][]{Hao2015}), some of which display phenotypic
plasticity (e.g., {\it Escherichia coli}, {\it Saccaromyces cerevisiae},
reviewed in \citep[][]{Chevin2012}).

Quantitative predictions of persistence for populations currently undergoing
ER could aid conservation planning, assessment of invasive species, and management
of antibiotic or pesticide resistance.  Although many examples of the latter
are major-gene effects (for analysis of ER via a single gene, see e.g.,
\citep[][]{Martin2013a}),
even these cases may include a quantitative contribution from minor genes \citep{Gomulkiewicz2010}.
Our theory is relevant for applied contexts where plasticity is thought to aid
persistence or invasion, including reintroduction for conservation purposes and
invasive species control.
For example invasive species are more plastic in response to added
resources \citep{Davidson2011}, suggesting more plastic species are better
invaders.  Our findings imply a more subtle prediction: a
``filter'' against invaders long-adapted to
low-reliability cues (where the same cue is maintained from the native to invaded
range), and a ``shield'' for regions where cues used by common invaders are
unreliable.
Applying this theory to a variety of systems might help resolve observed
variation how plasticity changes following a disturbance \citep[e.g.,][]{Crispo2010}.

\subsection{Conclusion}

Overall, evolving plasticity facilitates evolutionary rescue unless the new
environment is unpredictable. If it is not, then large variance in plasticity
might help lineages long-evolved to low-predictability environments adjust to
novel environments with high predictability.
These findings suggest the role of plasticity in longer-term evolution to
changing environments is positive, but limited.
The rapid increase in plasticity that occurs in our model
is an example of the Baldwin effect \citep{Lande2009} where plasticity increases
in species colonizing stressful environments.
(Baldwin actually proposed theory for the evolution of
plasticity that is much more general than this \citep{scheiner2014baldwin}.)
This effect, and related processes, have often been mentioned as
under-appreciated factors in evolution \citep{West-Eberhard2003,Laland2014}.
In novel environments that fluctuate with low predictability, however, we
show that a transient increase in plasticity (Figure~\ref{Fexplain}{\it (b)}) can
impose a substantial load on average growth, and thus a barrier to ER.
For populations whose plasticity evolved in response to
low-predictability cues, then, the Baldwin effect (as embodied in the two-phase
process studied here where plasticity increases) may have limited importance in adaptation.
An implication of these results is that over long timescales where the
environment has shifted frequently, we expect phenotypic plasticity (and its
genetic variance) to be absent or strongly reduced, unless cues are
consistently reliable.

\textbf{Code}
Code used to perform the simulations is available \citep{Ashander2016}.

\textbf{Competing interests}
We have no competing interests.

\textbf{Author contributions}
JA designed the study, carried out the analyses, and drafted the manuscript; LMC
designed the study, guided the analyses, and helped write the manuscript;
MLB designed the study, guided the analyses, and helped write the manuscript.
All authors gave final approval for publication.

\textbf{Acknowledgements}
Feedback from Swati Patel, Sebastian Schreiber, and Michael Turelli improved an
earlier version of the manuscript.

\textbf{Funding}
Support from IGERT (JA, NSF DGE-0801430 to P.I. Strauss), ContempEvol
(LMC, ANR-11-PDOC-005-01), and FluctEvol (LMC, ERC-2015-STG-678140-FluctEvol).

\section{References}

\begingroup
	\renewcommand{\section}[2]{}	\bibliography{refs}

\begin{thebibliography}{10}

\bibitem{Palumbi2001}
Palumbi S.
\newblock Humans as the world's greatest evolutionary force.
\newblock Science. 2001;293(5563):1786--1790.
\newblock http://dx.doi.org/10.1126/science.293.5536.1786.

\bibitem{Davis2005}
Davis MB, {Shaw} RG, {Etterson} JR.
\newblock Evolutionary responses to changing climate.
\newblock Ecology. 2005;86(7):1704--1714.
\newblock http://dx.doi.org/10.1890/03-0788.

\bibitem{Gomulkiewicz1995}
Gomulkiewicz R, {Holt} R.
\newblock When does evolution by natural selection prevent extinction?
\newblock Evolution. 1995;41(1):201--207.
\newblock \url{http://www.jstor.org/stable/2410305}.

\bibitem{Carlson2014}
Carlson SM, {Cunningham} CJ, {Westley} PAH.
\newblock Evolutionary rescue in a changing world.
\newblock Trends in {Ecology} \& {Evolution}. 2014;29(9):521--530.
\newblock http://dx.doi.org/10.1016/j.tree.2014.06.005.

\bibitem{deJong2005}
{de Jong} G.
\newblock Evolution of phenotypic plasticity: patterns of plasticity and the
  emergence of ecotypes.
\newblock The {New} {Phytologist}. 2005;166(1):101--17.
\newblock http://dx.doi.org/10.1111/j.1469-8137.2005.01322.x.

\bibitem{Crispo2007}
Crispo E.
\newblock The {Baldwin} effect and genetic assimilation: revisiting two
  mechanisms of evolutionary change mediated by phenotypic plasticity.
\newblock Evolution. 2007;61(11):2469--79.
\newblock http://dx.doi.org/10.1111/j.1558-5646.2007.00203.x.

\bibitem{Ghalambor2007}
Ghalambor CK, {McKay} JK, {Carroll} SP, {Reznick} DN.
\newblock Adaptive versus non-adaptive phenotypic plasticity and the potential
  for contemporary adaptation in new environments.
\newblock Functional {Ecology}. 2007;21(3):394--407.
\newblock http://dx.doi.org/10.1111/j.1365-2435.2007.01283.x.

\bibitem{Via1985}
Via S, {Lande} R.
\newblock Genotype-Environment Interaction and the Evolution of Phenotypic
  Plasticity.
\newblock Evolution. 1985;39(3):505--522.
\newblock \url{http://www.jstor.org/stable/10.2307/2408649}.

\bibitem{Lande2009}
Lande R.
\newblock Adaptation to an extraordinary environment by evolution of phenotypic
  plasticity and genetic assimilation.
\newblock Journal of {Evolutionary} {Biology}. 2009;22(7):1435--46.
\newblock http://dx.doi.org/10.1111/j.1420-9101.2009.01754.x.

\bibitem{Hoffmann1999}
Hoffmann A, {Meril}{\"a} J.
\newblock Heritable variation and evolution under favourable and unfavourable
  conditions.
\newblock Trends in {Ecology} \& {Evolution}. 1999;14(3):96--101.
\newblock
  \url{http://www.sciencedirect.com/science/article/pii/S0169534799015955}.

\bibitem{Charmantier2005}
Charmantier A, {Garant} D.
\newblock Environmental quality and evolutionary potential: lessons from wild
  populations.
\newblock Proceedings of the {Royal} {Society} {B}{\textendash}{Biological}
  {Sciences}. 2005;272(1571):1415--25.
\newblock http://dx.doi.org/10.1098/rspb.2005.3117.

\bibitem{McGuigan2009}
{McGuigan} K, {Sgr}{\`o} CM.
\newblock Evolutionary consequences of cryptic genetic variation.
\newblock Trends in {Ecology} \& {Evolution}. 2009;24(6):305--11.
\newblock http://dx.doi.org/10.1016/j.tree.2009.02.001.

\bibitem{Willis2008}
Willis CG, {Ruhfel} B, {Primack} RB, {Miller}-{Rushing} AJ, {Davis} CC.
\newblock Phylogenetic patterns of species loss in {Thoreau}'s woods are driven
  by climate change.
\newblock Proceedings of the {National} {Academy} of {Sciences}.
  2008;105(44):17029--33.
\newblock http://dx.doi.org/10.1073/pnas.0806446105.

\bibitem{Merila2014}
Meril{\"a} J, {Hendry} AP.
\newblock Climate change, adaptation, and phenotypic plasticity: the problem
  and the evidence.
\newblock Evolutionary {Applications}. 2014;7(1):1--14.
\newblock http://dx.doi.org/10.1111/eva.12137.

\bibitem{Davidson2011}
Davidson AM, {Jennions} M, {Nicotra} AB.
\newblock Do invasive species show higher phenotypic plasticity than native
  species and, if so, is it adaptive? {A} meta-analysis.
\newblock Ecology {Letters}. 2011;14(4):419--431.
\newblock http://dx.doi.org/10.1111/j.1461-0248.2011.01596.x.

\bibitem{Chevin2010}
Chevin LM, {Lande} R.
\newblock When do adaptive plasticity and genetic evolution prevent extinction
  of a density-regulated population?
\newblock Evolution. 2010;64(4):1143--50.
\newblock http://dx.doi.org/10.1111/j.1558-5646.2009.00875.x.

\bibitem{Moran1992}
Moran N.
\newblock The evolutionary maintenance of alternative phenotypes.
\newblock American {Naturalist}. 1992;139(5):971--989.
\newblock \url{http://www.jstor.org/stable/10.2307/2462363}.

\bibitem{Gavrilets1993}
Gavrilets S, {Scheiner} SM.
\newblock The genetics of phenotypic plasticity {V}. {Evolution} of reaction
  norm shape.
\newblock Journal of {Evolutionary} {Biology}. 1993;6:31--48.
\newblock http://dx.doi.org/10.1046/j.1420-9101.1993.6010031.x.

\bibitem{Lewontin1969}
Lewontin RC, {Cohen} D.
\newblock On population growth in a randomly varying environment.
\newblock Proceedings of the {National} {Academy} of {Sciences}.
  1969;62(4):1056--60.
\newblock \url{http://www.jstor.org/stable/59357}.

\bibitem{Lande2003}
Lande R, {Engen} S, {S}{\ae}ther BE.
\newblock Stochastic {Population} {Dynamics} in {Ecology} and {Conservation}.
\newblock Oxford {University} {Press}; 2003.

\bibitem{Turelli1977}
Turelli M.
\newblock Random environments and stochastic calculus.
\newblock Theoretical {Population} {Biology}. 1977;78:140--178.
\newblock \url{http://www.ncbi.nlm.nih.gov/pubmed/929455}.

\bibitem{Lande1996}
Lande R, {Shannon} S.
\newblock The {Role} of {Genetic} {Variation} in {Adaptation} and {Population}
  {Persistence} in a {Changing} {Environment}.
\newblock Evolution. 1996;50(1):434--437.
\newblock \url{http://www.jstor.org/stable/2410812}.

\bibitem{Burger1995}
{B}{\"u}rger R, {Lynch} M.
\newblock Evolution and Extinction in a Changing Environment : A
  Quantitative-Genetic Analysis.
\newblock Evolution. 1995;49(1):151--163.
\newblock \url{http://www.jstor.org/stable/2410301}.

\bibitem{Chevin2013a}
Chevin LM, {Collins} S, {Lef}{\`e}vre F.
\newblock Phenotypic plasticity and evolutionary demographic responses to
  climate change: taking theory out to the field.
\newblock Functional {Ecology}. 2013;27(4):967--979.
\newblock http://dx.doi.org/10.1111/j.1365-2435.2012.02043.x.

\bibitem{Reed2010}
Reed TE, {Waples} RS, {Schindler} DE, {Hard} JJ, {Kinnison} MT.
\newblock Phenotypic plasticity and population viability: the importance of
  environmental predictability.
\newblock Proceedings of the {Royal} {Society} {B}. 2010;227(1699):3391--3400.
\newblock http://dx.doi.org/10.1098/rspb.2010.0771.

\bibitem{Bjorklund2009}
Bj{\"o}rklund M, {Ranta} E, {Kaitala} V, a~{Bach} L, {Lundberg} P, {Stenseth}
  NC.
\newblock Quantitative trait evolution and environmental change.
\newblock {PloS} one. 2009;4(2):e4521.
\newblock http://dx.doi.org/10.1371/journal.pone.0004521.

\bibitem{Lande1976a}
Lande R.
\newblock Natural selection and random genetic drift in phenotypic evolution.
\newblock Evolution. 1976;30(2):314--334.
\newblock \url{http://www.jstor.org/stable/2407703}.

\bibitem{Tufto2015}
Tufto J.
\newblock Genetic evolution, plasticity, and bet-hedging as adaptive responses
  to temporally autocorrelated fluctuating selection: {A} quantitative genetic
  model.
\newblock Evolution. 2015;69(8):2034--2049.
\newblock http://dx.doi.org/10.1111/evo.12716.

\bibitem{MaynardSmith1976}
Maynard~{Smith} J.
\newblock What Determines the Rate of Evolution?
\newblock The {American} {Naturalist}. 1976;110(973):331--338.
\newblock \url{http://www.jstor.org/stable/2459757}.

\bibitem{deJong1999}
{de Jong} G.
\newblock Unpredictable selection in a structured population leads to local
  genetic differentiation in evolved reaction norms.
\newblock Journal of {Evolutionary} {Biology}. 1999;12(5):839--851.
\newblock http://dx.doi.org/10.1046/j.1420-9101.1999.00118.x.

\bibitem{RCoreTeam2014}
{R Core Team}.
\newblock {R}: {A} language and environment for statistical computing.
\newblock Vienna, {Austria}: {R} {Foundation} for {Statistical} {Computing};
  2014.
\newblock \url{http://www.r-project.org/}.

\bibitem{Eddelbuettel2011}
Eddelbuettel D, {Francois} R.
\newblock Rcpp: {Seamless} {R} and {C++} {Integration}.
\newblock Journal of {Statistical} {Software}. 2011;40(8):1--18.
\newblock \url{http://www.jstatsoft.org/v40/i08/}.

\bibitem{Ashander2016}
{Ashander} J, {Chevin} LM. phenoecosim v0.2.4; 2015.
\newblock ZENODO.
\newblock http://dx.doi.org/10.5281/zenodo.33933.

\bibitem{Caswell2001}
Caswell H.
\newblock Matrix {Population} {Models}: {Construction} {Analysis} and
  {Interpretation}.
\newblock 2nd ed. Sunderland, {MA}, {USA}: Sinauer {Associates}; 2001.

\bibitem{Michel2014}
Michel M, {Chevin} L, {Knouft} J.
\newblock Evolution of phenotype-environment associations by genetic responses
  to selection and phenotypic plasticity in a temporally autocorrelated
  environment.
\newblock Evolution. 2014;68(5):1374--1384.
\newblock http://dx.doi.org/10.1111/evo.12371.

\bibitem{Botero2015}
Botero CA, Weissing FJ, Wright J, Rubenstein DR.
\newblock Evolutionary tipping points in the capacity to adapt to environmental
  change.
\newblock Proceedings of the National Academy of Sciences.
  2015;112(1):184--189.
\newblock http://dx.doi.org/10.1073/pnas.1408589111.

\bibitem{Martin2013a}
Martin G, {Aguil}{\'e}e R, {Ramsayer} J, {Kaltz} O, {Ronce} O.
\newblock The probability of evolutionary rescue: towards a quantitative
  comparison between theory and evolution experiments.
\newblock Philosophical {Transactions} of the {Royal} {Society} of {London}
  {Series} {B}: {Biological} {Sciences}. 2013;368(1610):20120088.
\newblock http://dx.doi.org/10.1098/rstb.2012.0088.

\bibitem{Turelli1994}
Turelli M, {Barton} NH.
\newblock Genetic and Statistical Analyses of Strong Selection on Polygenic
  Traits: What, Me Normal?
\newblock Genetics. 1994;138(3):913--941.
\newblock http://dx.doi.org/10.1111/j.1558-5646.2009.00875.x.

\bibitem{Burgess2013}
Burgess SC, {Waples} RS, {Baskett} ML.
\newblock Local Adaptation When Competition Depends on Phenotypic Similarity.
\newblock Evolution. 2013;67(10):3012--3022.
\newblock http://dx.doi.org/10.1111/evo.12176.

\bibitem{Gomulkiewicz2010}
Gomulkiewicz R, {Holt} RD, {Barfield} M, {Nuismer} SL.
\newblock Genetics, adaptation, and invasion in harsh environments.
\newblock Evolutionary {Applications}. 2010;3(2):97--108.
\newblock http://dx.doi.org/10.1111/j.1752-4571.2009.00117.x.

\bibitem{deJong2000}
{de Jong} G, {Gavrilets} S.
\newblock Maintenance of genetic variation in phenotypic plasticity: the role
  of environmental variation.
\newblock Genetical {Research}. 2000;76(3):295--304.
\newblock \url{http://www.ncbi.nlm.nih.gov/pubmed/11204976}.

\bibitem{Kawecki2000}
Kawecki TJ.
\newblock The evolution of genetic canalization under fluctuating selection.
\newblock Evolution. 2000;54(1):1--12.
\newblock http://dx.doi.org/10.1554/0014-3820(2000)054.

\bibitem{Duputie2015}
Duputi{\'e} A, Rutschmann A, Ronce O, Chuine I.
\newblock Phenological plasticity will not help all species adapt to climate
  change.
\newblock Global {Change} {Biology}. 2015;21(8):3062--3073.
\newblock http://dx.doi.org/10.1111/gcb.12914.

\bibitem{Gomulkiewicz1992}
Gomulkiewicz R, {Kirkpatrick} M.
\newblock Quantitative genetics and the evolution of reaction norms.
\newblock Evolution. 1992;46(2):390--411.
\newblock \url{http://www.jstor.org/stable/2409860}.

\bibitem{Husby2011}
Husby A, {Visser} ME, {Kruuk} LEB.
\newblock Speeding up microevolution: the effects of increasing temperature on
  selection and genetic variance in a wild bird population.
\newblock {PLoS} {Biology}. 2011;9(2):e1000585.
\newblock http://dx.doi.org/10.1371/journal.pbio.1000585.

\bibitem{McGuigan2011}
{McGuigan} K, {Nishimura} N, {Currey} M, {Hurwit} D, {Cresko} WA.
\newblock Cryptic genetic variation and body size evolution in threespine
  stickleback.
\newblock Evolution; international journal of organic evolution.
  2011;65(4):1203--11.
\newblock http://dx.doi.org/10.1111/j.1558-5646.2010.01195.x.

\bibitem{Vedder2013}
Vedder O, {Bouwhuis} S, {Sheldon} BC.
\newblock Quantitative assessment of the importance of phenotypic plasticity in
  adaptation to climate change in wild bird populations.
\newblock {PLoS} {Biology}. 2013;11(7):e1001605.
\newblock http://dx.doi.org/10.1371/journal.pbio.1001605.

\bibitem{Gienapp2013}
Gienapp P, {Lof} M.
\newblock Predicting demographically sustainable rates of adaptation : can
  great tit breeding time keep pace with climate change?
\newblock Philosophical {Transactions} of the {Royal} {Society} {B}:
  {Biological} {Sciences}. 2013;368(1610):20120289.
\newblock
  \url{http://rstb.royalsocietypublishing.org/content/368/1610/20120289.short}.

\bibitem{Chevin2015}
Chevin LM, {Visser} ME, {Tufto} J.
\newblock Estimating the variation, autocorrelation, and environmental
  sensitivity of phenotypic selection.
\newblock Evolution. 2015;69(9):2319--2332.
\newblock http://dx.doi.org/10.1111/evo.12741.

\bibitem{Marshall2015}
Marshall DJ, {Burgess} SC.
\newblock Deconstructing environmental predictability: seasonality,
  environmental colour and the biogeography of marine life histories.
\newblock Ecology {Letters}. 2015;18(2):174--181.
\newblock http://dx.doi.org/10.1111/ele.12402.

\bibitem{Parmesan2006}
Parmesan C.
\newblock Ecological and {Evolutionary} {Responses} to {Recent} {Climate}
  {Change}.
\newblock Annual {Review} of {Ecology}, {Evolution}, and {Systematics}.
  2006;37(1):637--669.
\newblock http://dx.doi.org/10.1146/annurev.ecolsys.37.091305.110100.

\bibitem{Inouye2013}
Inouye DW, {Wielgolaski} FE.
\newblock Phenology: {An} {Integrative} {Environmental} {Science}.
\newblock In: Schwartz MD, editor. Phenology: {An} {Integrative}
  {Environmental} {Science}. vol. Second {Edition}. New {York}, {NY}, {USA}:
  Springer; 2013. p. 249--272.
\newblock http://dx.doi.org/10.1007/978-94-007-6925-0.

\bibitem{Anderson2012a}
Anderson JT, {Inouye} DW, {McKinney} AM, {Colautti} RI, {Mitchell}-{Olds} T.
\newblock Phenotypic plasticity and adaptive evolution contribute to advancing
  flowering phenology in response to climate change.
\newblock Proceedings of the {Royal} {Society} {B}: {Biological} {Sciences}.
  2012;1743(279):3843--52.
\newblock http://dx.doi.org/10.1098/rspb.2012.1051.

\bibitem{Bradshaw2008}
Bradshaw WE, {Holzapfel} CM.
\newblock Genetic response to rapid climate change: it's seasonal timing that
  matters.
\newblock Molecular {Ecology}. 2008;17(1):157--66.
\newblock http://dx.doi.org/10.1111/j.1365-294X.2007.03509.x.

\bibitem{Lee2011}
Lee CE, {Kiergaard} M, {Gelembiuk} GW, {Eads} BD, {Posavi} M.
\newblock Pumping ions: rapid parallel evolution of ionic regulation following
  habitat invasions.
\newblock Evolution. 2011;65(8):2229--44.
\newblock http://dx.doi.org/10.1111/j.1558-5646.2011.01308.x.

\bibitem{Hao2015}
Hao YQ, {Brockhurst} MA, {Petchey} OL, {Zhang} QG.
\newblock Evolutionary rescue can be impeded by temporary environmental
  amelioration.
\newblock Ecology {Letters}. 2015;18(9):892--898.
\newblock http://dx.doi.org/10.1111/ele.12465.

\bibitem{Chevin2012}
Chevin L, {Gallet} R, {Gomulkiewicz} R, {Holt} RD, {Fellous} S.
\newblock Phenotypic plasticity in evolutionary rescue experiments.
\newblock Philosophical {Transactions} of the {Royal} {Society} {B}:
  {Biological} {Sciences}. 2012;368:20120089.
\newblock http://dx.doi.org/10.1098/rstb.2012.0089.

\bibitem{Crispo2010}
Crispo E, DiBattista JD, Correa C, Thibert-Plante X, McKellar AE, Schwartz AK,
  et~al.
\newblock The evolution of phenotypic plasticity in response to anthropogenic
  disturbance.
\newblock Evolutionary Ecology Research. 2010;12(1):47--66.
\newblock
  \url{http://citeseerx.ist.psu.edu/viewdoc/summary?doi=10.1.1.184.1691}.

\bibitem{scheiner2014baldwin}
Scheiner SM.
\newblock The Baldwin effect: neglected and misunderstood.
\newblock The American Naturalist. 2014;184(4).
\newblock http://dx.doi.org/10.1086/677944.

\bibitem{West-Eberhard2003}
West-{Eberhard} MJ.
\newblock {Developmental Plasticity and Evolution}.
\newblock New {York}, {NY}, {USA}: Oxford {University} {Press}; 2003.

\bibitem{Laland2014}
Laland K, {Uller} T, {Feldman} M, {Sterelny} K, {M}{\"u}ller GB, {Moczek} A,
  et~al.
\newblock Does evolutionary theory need a rethink?
\newblock Nature. 2014;514(7521):161--164.
\newblock http://dx.doi.org/10.1038/514161a.

\end{thebibliography}
\endgroup

\newpage
\section{Figures}

\newpage
\appendix
\renewcommand*{\thefigure}{S\arabic{figure}}
\renewcommand*{\thetable}{S\arabic{table}}
\setcounter{figure}{0}
\setcounter{table}{0}
\setcounter{page}{1}
\begin{center}
	\textbf{Supplemental Information for \\``Predicting evolutionary rescue
		via evolving plasticity in stochastic environments''}

Jaime Ashander, Luis-Miguel Chevin, Marissa L. Baskett
\end{center}

\section{Approximate dynamics of growth rate in terms of maladaptation
	$x(t)$}\label{A:approxdyn}

Our equation~\eqref{Er} shows the Malthusian growth rate $r(t)$
is the sum of the maximum
Malthusian growth rate $r_{\max}$, the variance load $L_G$, and the lag load
$L_L$ with
	$r_{\max} = \log W_{\max}$,
	$L_G = \log \left ( \sqrt{S(\varepsilon_c(t)) \omega^2}\right)$, and
	$L_L = \frac{S(\varepsilon_c(t))}{2} x (t)^2$,
\begin{equation}\label{AElogW}
	\log \bar W(t) = r_{\max} + L_G(t) + L_L(t).
\end{equation}
Below, we derive an approximation to this where the only time dependence is
through the squared maladaptation $x(t)^2 = (\bar z(t)  -
\theta(\varepsilon_s(t)))^2$.

\subsection{Strength of selection}\label{A:ES}

The strength of selection, $S(\varepsilon_c(t))$, appears in both
the variance load $L_G$ and lag load $L_L$ but depends inversely on the phenotypic variance, which changes due to
stochastic variation in the environmental cue, which affects the phenotype.
Therefore, we approximate the expectation of the strength of selection using a Taylor series:
\begin{align*}
	\mathbb E [S (\varepsilon_c(t))] &=
	\left ( \mathbb E \left[ \sigma_z^2 (\varepsilon_c(t)) \right] + \omega^2 \right )^{-1} -
	\frac{\rm Var(\sigma_z^2)}{2\mathbb E[\sigma_z^2 ]^3} + O(\frac{\rm Var(\sigma_z^2)^3}{\mathbb E[\sigma_z^2 ]^4})\\
	&\approx \left ( \mathbb E \left[ \sigma_z^2 (\varepsilon_c(t)) \right]
		+ \omega^2 \right )^{-1}.
\end{align*}
Expanding $\sigma_z^2$ using equation~\eqref{Ez} of the main text and taking
expectations, we get the approximate expected strength of selection
${S}_\delta$, assuming $\delta \ll \sigma^2$, which we denote
\begin{equation}\label{ESdelta}
{S}_\delta =
\mathbb E [S (\varepsilon_c(t))]
\approx  \left ( \sigma_a^2 + \sigma_b^2 (\delta^2 + \sigma_c^2) + \sigma_e^2 + \omega^2 \right)^{-1}.
\end{equation}
Where we use the $\mathbb E [\varepsilon_c(t)^2] = {\rm Var}
[\varepsilon_c(t)] + \mathbb E [\varepsilon_c]^2$ and the
mean environmental shift $\delta$.

\subsection{Variance load}\label{A:ELG}

The first term of equation~\eqref{AElogW} is the phenotypic variance load at
time $t$
\[
	L_G(t) = 1/2 \log S (\varepsilon_c(t)) + 1/2 \log \omega^2.
\]
Taking expectations, and using Taylor series:
\begin{align*}
\mathbb E [L_G] &= 1/2 \log \omega^2 + 1/2 \mathbb E \left [ \log S (\varepsilon_c(t)) \right ]\\
	&\approx 1/2 \log \omega^2 + 1/2 \log \left (\mathbb E \left [ S (\varepsilon_c(t)) \right ] \right)
	- \frac{\rm Var(\sigma_z^2)}{2\mathbb E[\sigma_z^2 ]^2}\\
	&\approx 1/2 \log \omega^2 + 1/2 \log \left ( \mathbb E \left[ S
			(\varepsilon_c(t)) \right] \right )\\
	&\approx 1/2 \log \omega^2 + 1/2 \log  {S}_\delta
\end{align*}
where the right hand side of the last line is the approximate expected variance load
 after using $\mathbb E \left[S (\varepsilon_c(t)) \right] \approx {S}_\delta$
from~\eqref{ESdelta}.

\subsection{Lag load}\label{A:ELlag}

The second term of equation~\eqref{AElogW} is the lag load at time
$t$, $L_L(t)$. Here we show that in expectation, this load has two components, a
stochastic load and a shift load.  To derive expressions for
these, we take the expectation of the entire second term:
\begin{align*}
\mathbb E [ L_L(t)] =&\mathbb E \left[\frac{S (\varepsilon_c(t))}{2} x(t)^2 \right],\\
	=& \mathbb E \left [\frac{S (\varepsilon_c(t))}{2} \right]
\mathbb E \left [ x(t)^2 \right ] + {\rm Cov} \left [ \frac{S
		(\varepsilon_c(t))}{2}, x(t)^2 \right ],
\end{align*}
where we used the identity $\textrm{Cov } [ab] = \mathbb E[ab] - \mathbb E[a]\mathbb E [b]$.
The covariance in the final line
represents how stochastic changes in environment (and optimum) that cause
maladaptation also cause either larger or smaller phenotypic variance,
depending on the direction,  due to our assumption that genetic variance in plasticity
increases with $\delta$.  Furthermore, under weak stabilizing
selection, variance in $z$ contributes little to the strength of selection $S$.
For both these reasons, we expect this covariance to be small.

If we neglect the covariance term and use the approximate variance
load from~\ref{A:ELG} (${S}_\delta$ in main text) the expectation is
$\mathbb E [ L_L(t)] \approx \frac{{S}_\delta }{2} \mathbb E \left [ x(t)^2  \right ]$.
Removing the expectation, we use this as to approximate the dynamics of the lag
load
\begin{equation*}
L_L(t) = \frac{{S}_\delta }{2}  x(t)^2.
\end{equation*}

\subsection{Full dynamics}

Bringing together the approximations developed above, we have an expression for
the growth rate, where all components except the maladaptation are averaged over
the fluctuations,
\begin{equation}\label{Epopdyn1}
	r(t) \approx r_{\max} + 1/2 \log(\omega^2 {S}_\delta) -
	\frac{{S}_\delta}{2} x(t)^2.
\end{equation}
From this, we can obtain both the average long-run growth rate, by taking an
expectation to obtain $\bar r(t) \approx r_{\max} + 1/2 \log(\omega^2 
{S}_\delta) - \frac{{S}_\delta}{2} (\bar x(t)^2 + \sigma_x^2(t))$ (because
$\mathbb E [x^2] =  \bar x ^2 + \sigma_x^2$). This shows the lag load consists
of two loads.
The first, expected load from reduction in log mean fitness due to
maladaptation of the mean trait relative to the mean optimum in the new
environment, is ``shift load'' $- \frac{{S}_\delta}{2} \bar x(t)^2$. The
second, expected load due to reduction in log mean fitness due to random
fluctuations in the environment, is ``stochastic load''
$- \frac{{S}_\delta}{2} \sigma_x^2(t)$.  We can also analyse the variance in
trajectories of $r(t)$, and thus population growth because $\log N(t+1) = r(t) +
\log N(t)$.

Both of our subsequent analyses require computing the dynamics of mean squared
maladaptation $\bar x(t)^2$ and the variance in maladaptation $\sigma_x^2(t)$.

\section{Dynamics of shift load depend on mean maladaptation $\bar x(t)^2$}\label{A:Ez}

In this section, we derive the dynamics of the shift load
$- \frac{{S}_\delta}{2} \bar x(t)^2$ under the approximation
introduced by Lande \citep{Lande2009} that separates adaptation into a fast Phase 1
and a slow Phase 2. We demonstrate that the population is approximately
perfectly adapted in mean trait value by the end of Phase 1.  We first write
down the mean trait dynamics without the approximation, then describe the
timescales of the two phase approximation, and derive approximate dynamics of
the shift load owing to maladaptation in the mean trait.

\subsection{Mean trait}
Trait dynamics follow the standard equation $\Delta \mathbf y = \mathbf G
\boldsymbol \beta$, where $\mathbf y = (\bar a, \bar b)^T$, $\mathbf G$ is the additive genetic
variance-covariance matrix, and $\boldsymbol \beta$ is the selection
gradient.  The selection gradient on reaction norm height and slope
obtained by taking the log-gradient of $\bar W$ \citep[from
the equation for fitness above eq.~\eqref{Er} of main text;][]{Lande2009} is
\begin{equation}\label{AEbeta}
\boldsymbol \beta = - S(\varepsilon_c(t)) \left ( \begin{matrix} \bar a(t) - A + \bar b(t) \varepsilon_c - B \epsilon_s \\  \left(\bar a(t) - A + \bar b(t) \varepsilon_c - B \epsilon_s \right)\varepsilon_c \end{matrix} \right ).
\end{equation}
With a constant additive genetic variance-covariance matrix ($\mathbf
G$ matrix), the change per generation in $\bar a(t)$ and $\bar b(t)$
is given by
\[
\Delta \left( \begin{matrix} \bar a(t)\\ \bar b(t)\end{matrix} \right) = \left ( \begin{matrix} \sigma_a^2 &0 \\ 0 &\sigma_b^2 \end{matrix} \right)  \boldsymbol \beta.
\]
In the new environment, the expectation of the change per generation conditional
on $\bar a$ and $\bar b$ is
\begin{align*}
\Delta \left( \begin{matrix} \mathbb E(\bar a(t)) \\\mathbb E(\bar b(t))  \end{matrix}\right) &= \mathbb E [ \mathbf G \boldsymbol \beta ]\\
&\approx
-{S}_\delta
\mathbf G
\left( \begin{matrix}
	\bar a - A + \bar b \delta - B \delta \\
	\mathbb E[(\bar a - A)\varepsilon_c(t)] + \mathbb E[\bar b \varepsilon_c^2(t)]  -B \mathbb E[  \varepsilon_c \varepsilon_s]
	\end{matrix}
\right),
\end{align*}
where the approximation comes from the treating $\mathbb E
[S(\varepsilon_c(t))]$ as a constant.  We assume no environmental
tracking by phenotypic plasticity, i.e., ${\rm Cov} [ \bar b_t,
\varepsilon_c(t)^2(t) ] = 0$ so $\mathbb E[ \bar b_t \varepsilon_c^2 ] =
\mathbb E[ \bar b_t]\mathbb E[ \varepsilon_c(t)^2(t) ]$, or by reaction norm
elevation, i.e., ${\rm Cov} [ \bar a_t, \varepsilon_c(t) ] = 0$ so
$\mathbb E[ \bar a_t, \varepsilon_c(t) ] = \mathbb E[ \bar a_t]\mathbb E[
\varepsilon_c(t)]$.  Note that the tracking of the environment by the
reaction norm elevation could be included, reducing the expected mean
plasticity \citep[][]{Tufto2015}, but we neglect it
here.  However, we do allow for environmental tracking when computing
the lag load (below).  Then, using $\mathbb E[\varepsilon^2(t)] =
\delta^2 + \sigma^2$ and $\mathbb E[\varepsilon_c(t)\varepsilon_s] =
\delta^2 + \rho \sigma^2$, the expectation of the change is approximately
\begin{equation}\label{AEexpchange}
\mathbb E \left [
	\Delta \left( \begin{matrix} \bar a(t) \\
	\bar b(t)  \end{matrix}\right)
\right]
\approx  -{S}_\delta
\mathbf G
\left [
	\left(\begin{matrix} 1 & \delta \\
\delta &\delta^2 \end{matrix}\right)
	\left(\begin{matrix} \bar a(t) - A \\
	\bar b(t) - B \end{matrix}\right)
	+
	\left(\begin{matrix} 0\\
	(\bar b(t) - \rho B) \sigma^2 \end{matrix}\right)
\right].
\end{equation}
The approximation is exact if ${\rm Cov} [ \bar b_t,
\varepsilon_c(t)^2(t) ] = {\rm Cov} [ \bar a_t, \varepsilon_c ] = 0$.
Note that this differs from Lande \citep{Lande2009}, where this relation
was treated as exact \citep[][]{Tufto2015}.

We solve for the explicit trait dynamics relative to the long-run
equilibrium state \citep{Lande2009}.  Setting selection gradient
$\boldsymbol \beta$ in equation~\eqref{AEbeta} to zero, solve for
long run trait values
\begin{equation}\label{AEzinf}
\left(\begin{matrix}
		\bar a_\infty \\
		\bar b_\infty
\end{matrix}\right)
=
\left(\begin{matrix}
		A+B \delta (1 - \rho_\delta)\\
		B \rho_\delta
\end{matrix}\right)
\end{equation}
One can show, with some algebra \citep{Lande2009}, the one-generation
change~\eqref{AEexpchange} is the product -${S}_\delta
\mathbf {\tilde G} \mathbf z(t)$, where $\mathbf z(t)$ is the
difference between mean trait values and their long-run equilibrium values
computed above and
\[
\mathbf {\tilde G} = \left(
	\begin{array}{cc}
		\sigma _a{}^2 & \sigma _a{}^2\delta  \\
		\sigma _b{}^2\delta  & \sigma _b{}^2\delta ^2\left( 1+ \tfrac{\sigma ^2}{\delta^2}\right) \\
	\end{array}
\right).
\]
The expected dynamics can then be expressed in terms of the
eigenvectors and eigenvalues of the matrix $\mathbf {\tilde G}$
\citep{Lande2009,Chevin2010},
\begin{equation}\label{AEdyn}
\mathbf z(t) = c_1 \mathbf e_1 \left(1-{S}_\delta  \lambda _1\right)^t+c_2 \mathbf e_2 \left(1-{S}_\delta  \lambda _2\right)^t,
\end{equation}
where $\mathbf e_i$ and $\lambda_i$ are eigenvectors and eigenvalues
of $\mathbf {\tilde G}$ respectively and $c_i$ terms are constants
determined by initial conditions.

\subsection{Approximation for a large environmental shift}

As in earlier work, we consider the case where the shift in the mean
environment is very large relative to background noise,
$\tfrac{\sigma ^2}{\delta^2} \ll 1$.  Here, we initially write down
the eigenvalues and eigenvectors of $\mathbf {\tilde G}$ to first
order in this small term, but thereafter follow
Chevin and Lande \citep{Chevin2010} in deriving approximate dynamics to leading
order. To first order in  $\tfrac{\sigma ^2}{\delta^2}$ , the eigenvalues are
\[
\left(
	\begin{matrix}
		\lambda _1 \\
		\lambda _2
	\end{matrix}
\right)
=
\left(
	\begin{matrix}
		\left(\sigma _a^2+\delta ^2 \sigma _b^2\right)+\delta ^2 \sigma _b^2 \tfrac{\sigma ^2}{\delta^2} \phi \\
		\sigma _a^2 \tfrac{\sigma ^2}{\delta^2}\phi
	\end{matrix}
\right),
\]
and the eigenvectors are
\[
\mathbf e_1 = \left(
	\begin{array}{c}
		\delta  \left(1-\phi\right) \left(1-\tfrac{\sigma ^2}{\delta^2} \phi\right) \\
		\phi \\
	\end{array}
\right),
\quad
\mathbf e_2 = \left(
	\begin{array}{c}
		\delta  \left(1+\tfrac{\sigma ^2}{\delta^2}  \phi \right) \\
		-1 \\
	\end{array}
\right).
\]
These match the calculations of \citep{Lande2009} up to a constant of
${S}_\delta$ (equivalent to $\gamma$ in Lande \citep{Lande2009}).

Assuming the population has long evolved in an environment with
predictability $\rho$, the initial trait values are $(\bar a_0,
\bar b_0) = (A, \rho B)$.  Using~\eqref{AEzinf}, the initial
conditions in the re-centred trait $x$ are
\[
	\mathbf z_0 =
\left(
	\begin{array}{c}
		-B \delta (1 -  \rho_\delta) \\
		B (\rho - \rho_\delta) \\
	\end{array}
\right)
\]
To leading order, the constants are
\[
\left(
	\begin{array}{c}
		\text{c1} \\
		\text{c2} \\
	\end{array}
\right) = \left(
	\begin{array}{c}
		-B (1 -\rho ) \\
		-B(\rho (1 - \phi) - \rho_\delta + \phi) \\
	\end{array}
\right)
\]

\subsubsection{Timescales of phases 1 and 2}\label{A:timescales}
If most phenotypic variation in the new environment is due to variance
in plasticity, $\phi \approx 1$, and the shift in the mean environment
is large (relative to background variability as in our approximation
above), the trait change takes place in two phases that occur at very
different timescales \citep{Lande2009}.  When selection is weak,
geometric terms in~\eqref{AEdyn} can be replaced by exponential terms
$e^{-t {S}_\delta \lambda _i}$, indicating the relative
timescales of change along the eigenvectors $e_i$ are given by $t_i
\approx \tfrac{1}{\lambda_i}$.  The ratio $t_1/t_2 \approx \phi ( 1 -
\phi) \tfrac{\sigma^2}{\delta^2}$, and when much of the additive
genetic variation is due to variation in plasticity so $\phi \approx
1$, then $t_1/ t_2 \approx \tfrac{\sigma^2}{\delta^2}$, which is small
in the approximate case we treat. Then, change along $\mathbf e_1$ occurs
very fast relative to change along $\mathbf e_2$ \citep{Lande2009}.
At the end of Phase 1, $e^{-t {S}_\delta \lambda _1} \approx 0$
while $e^{-t {S}_\delta \lambda _2} \approx 1$ Thus, the
approximate state of the system relative to its final state, i.e.,
\eqref{AEzinf}, is $c_2 e_2$. The trait values,
at the end of Phase 1 are, to leading order,
\begin{equation}\label{AEz1}
\mathbb E \left [
	\left(
		\begin{matrix}
			\bar a_{O(t_1)}\\
			\bar b_{O(t_1)}
		\end{matrix}
	\right)
\right]
\approx
\left(
	\begin{array}{c}
		A+B\delta (1 - \rho ) (1 -\phi ) \\
		B (\rho +\phi (1 -\rho)) \\
	\end{array}
\right)
\end{equation}
The effect of the initial environment occurs through predictability
$\rho$, which under our assumption that the population is adapted
initially also determines the initial mean plasticity $\bar b_0 =
\rho B$. We see the initial plasticity has a strong influence at the
end of Phase 1 only if $\phi$ is small. When $\phi$ is large, the
plasticity at the end of Phase 1 is close to ``perfect''
i.e. $b_{O(t_1)} \approx B$. Note also that to first order, the mean
phenotype is perfectly adapted $\bar z_{O(t_1)} \approx A + B
\delta$.

Where the extinction risk is calculated at half of the characteristic timescale
of Phase 2, i.e.,  $t_{\rm bef} = \tfrac{\phi \delta^2 }{2 \sigma_a^2
	\sigma^2}$

\subsubsection{Trait dynamics during Phase 1}
Throughout Phase 1, the term $\left(1 -{S}_\delta \lambda
_2\right)^t \approx 1$, so the dynamics are given by
\[
\mathbf z(t) = c_2 \mathbf e_2+c_1 \mathbf e_1 \left( 1-{S}_\delta  \lambda_1\right)^t.
\]
We again replace the geometric term with an exponential (valid for
weak selection) and re-normalize.  To leading order, after some
rearranging, the right hand side of the expected dynamics is
\begin{equation}\label{AEtraitdyn}
\mathbb E
\left[
	\left (
	\begin{matrix}
		\bar a(t)\\
		\bar b(t)
	\end{matrix}
\right)
\right]
=  \left(1-e^{-t {S}_\delta  \lambda _1}\right) B (1-\rho )\left(
\begin{array}{c}
	\delta (1-\phi ) \\
	\phi  \\
\end{array}
\right)+ \left(
\begin{array}{c}
	A \\
	B \rho  \\
\end{array}
\right),
\end{equation}
which is analogous to the result of Chevin and Lande \citep[Supporting Information,
eq A6;][]{Chevin2010}.  As in that paper, we compute the
eigenvalue only to leading order in $\tfrac{\sigma ^2}{\delta^2}$ so
$\lambda_1 \approx \sigma _a^2+\delta ^2 \sigma _b^2$
which is equivalent to the expression $\frac{\sigma _a^2}{1-\phi }$
used in Chevin and Lande \citep{Chevin2010}.

\subsubsection{Trait dynamics during Phase 2}\label{Aphase2}

During Phase 2, the term $\left(1 -{S}_\delta \lambda
_1\right)^t \approx 0$, so the dynamics are given by
\[
\mathbf z(t) = c_2 \mathbf e_2 \left( 1 - {S}_\delta  \lambda_2\right)^t.
\]
Then, using~\eqref{AEzinf} and again replacing the geometric term
with an exponential (valid for weak selection) and re-normalizing, the
expected dynamics to leading order in $\tfrac{\sigma ^2}{\delta^2}$
during Phase 2 are
\begin{equation}\label{AEtraitdyn2}
\mathbb E
\left[
	\left (
	\begin{matrix}
		\bar a(t)\\
		\bar b(t)
	\end{matrix}
\right)
\right]
=
\left(\begin{matrix}
	A+B \delta (1 - \rho_\delta)\\
	B \rho_\delta
\end{matrix}\right)
-
B (\rho - \rho_\delta + \phi (1- \rho)) \\
\left(
\begin{array}{c}
	\delta   \\
	-1 \\
\end{array}
\right)
e^{-t {S}_\delta \sigma_a^2 \tfrac{\sigma ^2}{\delta^2} \phi},
\end{equation}
where we have used $\lambda_2 \approx \sigma_a^2 \tfrac{\sigma
^2}{\delta^2} \phi$.  Note that for $t = O(t_1)$, this exponential
term equals 1 and this equation agrees with~\eqref{AEz1}.

\subsection{Dynamics of expected maladaptation during Phase 1}\label{A:Emalad}
We derive the expected maladaptation of the mean trait during an initial phase
of evolutionary rescue, focusing on a case where the size of the environmental
shift is large and much of the additive genetic variance in the new environment
owes to genetic variance in reaction norm slope.  The shift load (computed
in~\ref{A:approxdyn}) is  $\frac{{S}_\delta}{2} \bar x(t)^2$.
After we compute the dynamics of the mean maladaptation $\bar x(t)^2$, we will have an
approximation for dynamics of the shift load,
\begin{align*}
	\bar x(t)^2 = \mathbb E[x(t)]^2 =&
	\left(\mathbb E[\bar a(t)] - A +   \mathbb E [\bar b(t) \varepsilon_c(t)] - B \mathbb E [\varepsilon_s] \right)^2\\
	\approx&  \left(\mathbb E[\bar a(t)] - A +   \delta (\mathbb E [\bar b(t)]  - B ) \right)^2.
\end{align*}
The last equation comes from assuming the covariance between reaction
norm slope and the cuing environment is small relative to the mean
value of the new environment.  Then, $\mathbb E [\bar b(t)
\varepsilon_c(t)] \approx \mathbb E [\bar b(t)] \mathbb E [
\varepsilon_c(t)] = \delta \mathbb E [\bar b(t)]$ in the new
environment.  This is reasonable when $\tfrac{\sigma ^2}{\delta^2}
\ll 1$.  After using~\eqref{AEtraitdyn} and some algebra, we obtain
  \begin{equation}\label{Eldfull}
	\bar x(t) ^2 \approx
	B^2 \delta^2 (1-\rho )^2  e^{- 2 t {S}_\delta \frac{\sigma _a^2}{1-\phi}}.
  \end{equation}  Where we use $\phi$ to represent the
  proportion of additive genetic variation in the new environment due
  to variation in plasticity,
  \[
	\phi = \frac{\delta ^2 \sigma _b^2}{\sigma _a^2+\delta ^2 \sigma _b^2}.
  \]
Equation~\eqref{Eldfull} indicates the shift load goes to zero as $t$ increases.

\section{Stochastic load: variance in maladaptation $\sigma_x^2$ at stationarity}\label{A:ELs}

\subsection{Perceived environment with fixed plasticity}

A tactic from Michel \textit{et al.} \citep{Michel2014} aids in calculating the variance as a function
of fixed plasticity.
We define the perceived optimum $\psi(t)$ as the
difference between the optimum and the mean trait after accounting for the
plastic response $\psi(t) = B\varepsilon_s(t) - \bar b^* \varepsilon_c(t)$ so
that $x(t) = \bar a(t) - \psi(t)$.
Then, the perceived variance in the optimum is
\begin{equation}\label{AEsigmapsi}
	\sigma_\psi^2(\bar b^*, \rho_\delta) =
	\sigma^2 ( B^2  + \bar b^* (\bar b^*  - 2 B \rho_\delta)),
\end{equation}
and autocorrelation in the perceived optimum is
\begin{equation}\label{AETpsi}
	T_\psi(\bar b^*, \rho_\delta) = -\log \left [ \rho_\delta^{1/\tau}
	\left ( 1 - \frac{B \bar b^* (\rho_\delta^{-1} - \rho_\delta)}
		{B ^2 + \bar b^*(\bar b^* - 2 B \rho_\delta)} \right
	) \right ]^{-1}
\end{equation} \citep{Michel2014}.
We can then express maladaptation in terms of the intercept and
perceived environment as $x(t) = \bar a(t) - \psi(t)$.

\subsection{Variance at stationarity}

We derive an approximation for variance in maladaptation under stationarity,
which we denote $\sigma_x^2$.
In practice, this means we solve for the effect of fluctuations on maladaptation
after a long time, we assume the mean
maladaptation is zero, and we also assume fixed mean plasticity $\bar b^*$.
We are interested in finding an asymptotic  expression for the variance of this term.

Assuming fixed plasticity, all change in the trait occurs through
change in the reaction norm height,
\[
	\Delta \bar z = \Delta \bar a(t) = - {S}_\delta \sigma^2_a x(t).
\]
When selection is weak relative to genetic variance in reaction norm
height, and the fluctuations in the perceived environment are not
large, evolution can be approximated in continuous time
\citep{Lande1996,Michel2014} as
\[
	\frac{d x}{d t} + {S}_\delta \sigma^2_a x = - \frac{d\psi}{dt},
\]
where $x= \bar a(t) - \psi$.  For $t \gg t_1$ and constant genetic
variance $\sigma_a^2$, the solution to this differential equation is
\begin{equation}\label{AEa}
	\bar a(t) = {S}_\delta \sigma^2_a \int_0^\infty \exp
	\left( - {S}_\delta \sigma^2_a \tau \right) \psi({t - \tau}) {\rm d} \tau.
\end{equation}
What remains is to compute ${\rm Var}[x(t)]$. Using our
quasi-stationarity assumption, we need
only compute $ \mathbb E[x(t)^2] =
- 2 \mathbb E[\bar a(t) \psi(t)]
+ \mathbb E[\bar a(t)^2] + \mathbb E[\psi(t)^2]$.
The last of these expectations is simply the variance of the
perceived environment $\sigma^2_\psi$.
The first and second expectations integrate over time (from eq.~\ref{AEa}),
\begin{align*}
	- 2 \mathbb E[\bar a(t) \psi(t)] &=  -2 {S}_\delta \sigma^2_a  \int_0^\infty \exp
	\left( - {S}_\delta \sigma^2_a \tau \right) \mathbb E[\psi(t)\psi({t - \tau})] {\rm d} \tau \\
	\mathbb E[\bar a(t)^2] &=    {S}_\delta^2 \sigma^4_a \int_0^\infty \int_0^\infty \exp
	\left( - {S}_\delta \sigma^2_a ( \tau_1 + \tau_2) \right) \mathbb E[\psi({t - \tau_1})\psi({t - \tau_2})] {\rm d} \tau_1 {\rm d} \tau_2.
\end{align*}
Because $\psi$ is a linear combination of autoregressive Gaussian
processes $\varepsilon_c(t)$ and $\varepsilon_s$, we can express
the expectations involving $\psi$ in terms of autocovariance
$\mathbb E[\psi(t) , \psi(t-\tau) ] = \sigma_\psi^2 \exp (-
\tau / T_\psi) $ and $\mathbb E[\psi(t-\tau_1) , \psi(t-\tau_2) ] =
\sigma_\psi^2 \exp (- |\tau_1 - \tau_2 | / T_\psi) $.  In both of
these expressions, $T_\psi$ is the characteristic autocorrelation time
of the perceived environment $\psi$.

The first expectation is
\begin{align*}
	- 2 \mathbb E[\bar a(t) \psi(t)] &=   - 2{S}_\delta \sigma^2_a  \sigma_\psi^2 \int_0^\infty \exp
	\left( - \tau ({S}_\delta \sigma^2_a  + 1  / T_\psi) \right) {\rm d} \tau, \\
		&= - 2 \frac{T_\psi{S}_\delta \sigma^2_a  \sigma_\psi^2 }{(T_\psi{S}_\delta \sigma^2_a  +  1 )}.
\end{align*}

The second expectation involves an absolute value term, meaning the integral must be taken in two parts, and
evaluates to
\begin{align*}
	\mathbb E[\bar a(t)^2] =& {S}_\delta^2 \sigma^4_a \int_0^\infty \int_0^\infty \exp
	\left( - {S}_\delta \sigma^2_a ( \tau_1 + \tau_2) \right) \sigma_\psi^2 \exp (- |\tau_1 - \tau_2 | / T_\psi) {\rm d} \tau_1 {\rm d} \tau_2,\\
	=& {S}_\delta \sigma^2_a  \sigma_\psi^2 \frac{T_\psi}{\left(T_{\psi} {S}_\delta  \sigma _a^2 + 1 \right)}
\end{align*}
Combining these expressions,
\begin{align*}
	\mathbb E[( \bar a(t) - \psi(t))^2] &=
	\sigma_\psi^2
	\left(
		- 2 \frac{T_\psi{S}_\delta \sigma^2_a   }{(T_\psi{S}_\delta \sigma^2_a  +  1 )}
		+ {S}_\delta \sigma^2_a   \frac{T_\psi}{\left(T_{\psi} {S}_\delta  \sigma _a^2 + 1 \right)}
	+ 1 \right)\\
	&=
	\frac{\sigma_\psi^2}{\left(T_{\psi} {S}_\delta  \sigma _a^2 + 1 \right)}
	\left(T_{\psi} {S}_\delta  \sigma _a^2 + 1
		- 2 T_\psi{S}_\delta \sigma^2_a
		+   T_\psi{S}_\delta \sigma^2_a
	\right)\\
	\sigma_x^2(\bar b ^ *, \delta, \rho_\delta)
	&:=
	\frac{\sigma_\psi^2}{\left(T_{\psi} {S}_\delta  \sigma _a^2 + 1 \right)},
\end{align*}
where the last line defines the variance we sought to calculate.
As our notation emphasizes, this load depends primarily on the level of
plasticity $\bar b ^ *$ and the size of the shift $\delta$.
Note that this formula matches Lande and Shannon \citep{Lande1996}.

\subsection{Quasi-stationary variance over Phase 1}\label{A:quasistat}

The variance in maladaptation $\sigma_x^2$ depends on variance $\sigma_\psi^2(\bar
b^*, \rho_\delta)$ from eq.~\eqref{AEsigmapsi} and characteristic timescale
$T_\psi(\bar b^*, \rho_\delta)$ from eq.~\eqref{AETpsi} of perceived
fluctuations in the perceived optimum (given autocorrelation $\rho_\delta$ at timescale
$\tau$ in the true optimum), as
\begin{equation}\label{ELs}
	\sigma_x^2(\bar b^*, \delta, \rho_\delta) =
	\frac{\sigma_\psi^2 (\bar b^*, \rho_\delta)}
	{\left( {S}_\delta  \sigma_a^2 T_\psi(\bar b^*, \rho_\delta) + 1 \right)}.
\end{equation}

After a long time, on the timescale of Phase 1, $t_1$, the mean maladaptation
is zero, as shown in eq.~\eqref{Eldfull}, and the variance in maladaptation is
stationary. At this point, as shown in~\ref{A:Ez}, plasticity is at its
approximate maximum $b_{\rm max} = B(\rho +\phi (1 -\rho))$.  It remains as this
value for a long time (on the scale of Phase 1; although eventually decays to
the predictability $\rho_\delta$ according to the dynamics in~\ref{Aphase2}) and
so we evaluate~\eqref{ELs} at $b_{\rm max}$ to determine persistence
in~\eqref{Eltpersist}. Using the ``quasi-stationary'' approximation that
fluctuations achieve stationarity while $\bar b ^ * = b_{\rm max}$, we evaluate
$\sigma_x^2(b_{\rm max}, \delta, \rho_\delta)$.

\begin{figure}[h]
	\begin{center}
		\includegraphics{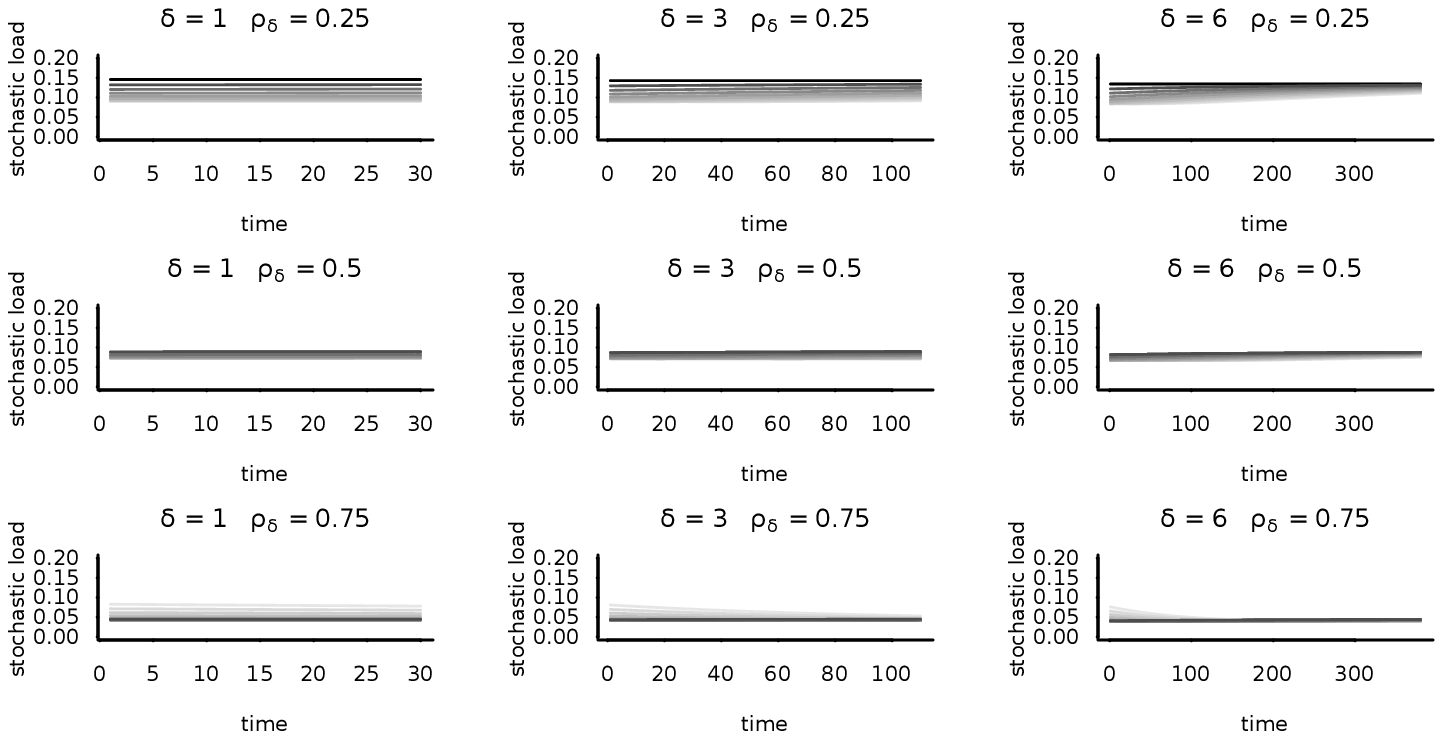}
	\caption[Quasi-stationary stochastic load]{Quasi-stationary stochastic
		load plotted against time during Phase 1 (up to the
		characteristic timescale of Phase 2 $t_2$) for the same values
		of relative plasticity $\alpha$ as Figure~\ref{FS2stochload}. While
		changes in the magnitude of stochastic load over Phase 1 are not
		large, and in some cases not apparent, when the change reduces
		(or increases) the mismatch the stochastic load follows (e.g.,
		bottom right or top right panel).  Other parameters are
		$\sigma_b^2=$0.05, $\tilde S(\delta) = $0.0446429, $B=$2,
		$\sigma^2=$ 1, $\sigma_a^2=$0.1.
	} \label{FS1quasistochload}
\end{center}
\end{figure}

Note that we could also evaluate~\eqref{ELs} at any time $t$ during Phase 1,
\begin{equation}\label{ELsqs}
	\sigma_x^2(t) = \sigma_x^2(\bar b(t), \delta, \rho_\delta),
\end{equation}
by using dynamics for reaction norm slope $\bar b(t)$ from eq.~\eqref{AEtraitdyn}.
Doing so assumes for each change in $\bar b$ over Phase 1 the stochastic
variance in maladaptation achieves stationarity. No doubt this is inaccurate
in some cases, but it is a tractable analytical approximation for the variance.
Using this approximation over the parameter ranges we examine in this paper, the
stochastic load does not change much with changes in plasticity
(Figure~\ref{FS1quasistochload}). Accordingly, we use the simpler approximation
above, and evaluate the variance at $b_{\rm max}$ for all time.

\section{Simulation}\label{A:envsim}
	In simulation, we implement the autocorrelated environment as an
	autoregressive function with correlation $\kappa$ on a timescale with $n =
	{\rm ceiling} [\frac{1}{\tau}]$ time units in a generation.  The
	process simulated is
	\[
		x_i = \kappa x_{i - 1} + \sqrt{ 1 - \kappa^2} \xi \sigma,
	\]
	where $\xi$ is a unit normal random variable.  If $x$ and $\xi$ are
	independent (as generally assumed in autoregressive functions) and the
	process is stationary, it has variance $\sigma^2$ (and mean $0$).
	Further, the covariance of observations one time unit apart is $\kappa
	\sigma^2$ and the correlation of such pairs of observations is
	$\kappa$. For observations $n$ time units apart, the correlation
	becomes $\kappa^n$.  Thus, using the $n$ time steps per generation, the
	simulated process relates to the exponential autocovariance function
	given above as $\kappa ^ n = \rho ^ {1 / \tau }$.  Accordingly we set the
	correlation within simulations equal to that of the
	environmental predictability at timescale $\tau$, i.e., $\kappa
	= \rho$.

\section{Alternative assumptions of reaction norm shape}\label{A:otherRN}

If, instead of assuming phenotypic variance in the reference environment is
minimal, we assume the environment is shifted to an environment where the
variance is minimal, then the additive genetic variance of the expressed trait
$z(t)$ increases quadratically away from the novel environment
$\varepsilon_c(t) = \delta$.
Without the assumption that variance is minimized in the reference environment,
the additive genetic covariance between $a$ and $b$ is non-zero
in the reference environment \citep{Lande2009}. From that paper, the full form of~\eqref{Ezvar} is
$\sigma_z^2(\varepsilon_c(t)) = \sigma^2_a + 2 \sigma_{ab} \varepsilon_c (t) +
\sigma^2_b \varepsilon_c^2(t) + \sigma_e^2$, where $\sigma_{ab}$ is the additive
genetic covariance between reaction norm slope and variance in the reference
environment.
(With the assumption of variance
minimized at $\varepsilon_c = 0$, $\sigma_{ab} = 0$, and the covariance in any
other environment is ${\rm Cov}(a, b) = \sigma_b^2 \varepsilon_c$.)
Then $\sigma_z^2(\varepsilon_c(t))$ is minimized in the environment
$\varepsilon_c^* = - \sigma_{ab} / \sigma_b^2$ \citep{Lande2009}.

Thus, assuming minimal variance in the new environment implies
$\sigma_{ab} =  - \delta \sigma_b^2$.
The mean of the expressed trait value $z(t)$
before selection is the same as in~\eqref{Ez}, but the variance differs, giving
\begin{subequations}\label{Ezdiff}
	\begin{align}
		&\bar z(t) = \bar a(t) + \bar b(t) \varepsilon_c(t)\\
  &\sigma_z^2(\varepsilon_c(t)) = \sigma^2_a - 2 \sigma_b^2 \delta \varepsilon_c (t) +  \sigma^2_b \varepsilon_c^2(t) + \sigma_e^2,
	\end{align}
\end{subequations}
which assumes the additive genetic variances are constant in time.

In this case, the expressed trait $z$ and
the slope $b$ have covariance ${\rm Cov}(z,b) = (\varepsilon_c(t) - \delta)
\sigma_b^2$ and so, with $\varepsilon_c(t) \approx \delta$  there is
approximately zero covariance between the trait and reaction norm slope, and so
direct selection on the trait results in very weak selection on reaction norm
slope.

Consequently, the transient increase in plasticity should not be expected without
the assumed increase of genetic variance in novel environments. Given the uncertainty
described by McGuigan and Sgro \citep{McGuigan2009} concerning the effects of stress (i.e.,
novelty) on additive genetic variance, theory could usefully outline empirical
possibilities.  The derivation of~\eqref{Ezdiff} is just the beginning. It does
show, however, that the theory presented in the main text is not completely general.
No theory is
unless it thoroughly considers the possible relationships between additive
genetic variance and environmental shifts. In the main text, we analyse only one
set---where many are possible---of assumptions on how plasticity, demography,
and evolution interact during evolutionary rescue.
However, this set of assumptions appears met by empirical reality in at least
some cases (see main text, "Assumptions and Caveats").

\section{A small shift and large additive variance in plasticity
	$\sigma_b^2$}\label{A:approxbad}

We expect our approximation to perform best when a large proportion of additive
variance in the novel environment is due to variance plasticity (i.e., $\phi
\approx 1$).  This occurs in our model either when $\sigma_b^2$ (GxE) is large
in the reference environment, or when a large shift in the mean environment
causes quadratic increases in additive genetic variance (as $\delta ^2
\sigma_b^2$).

In the former case, the population has inherently high genetic variability in
plasticity, and our assumptions imply a relatively large changes in additive
genetic variance for even small shifts in the mean environment.
In the latter case,  increased additive genetic variance is driven by the
novelty of the environment.

Simulations reveal that despite both cases increasing $\phi$, the approximation
does not perform equivalently well. If the mean shift is small then the
approximation for the threshold between decline and persistence performs poorly
as additive variance in plasticity increases ($\delta = 1.5$, Figure~\ref{FS6potrescuesmalldelta}).
In fact, the approximation appears to perform better for lower values of $\phi$.

\section{Supplementary Figures}

\begin{figure}[h]
	\begin{center}
		\includegraphics{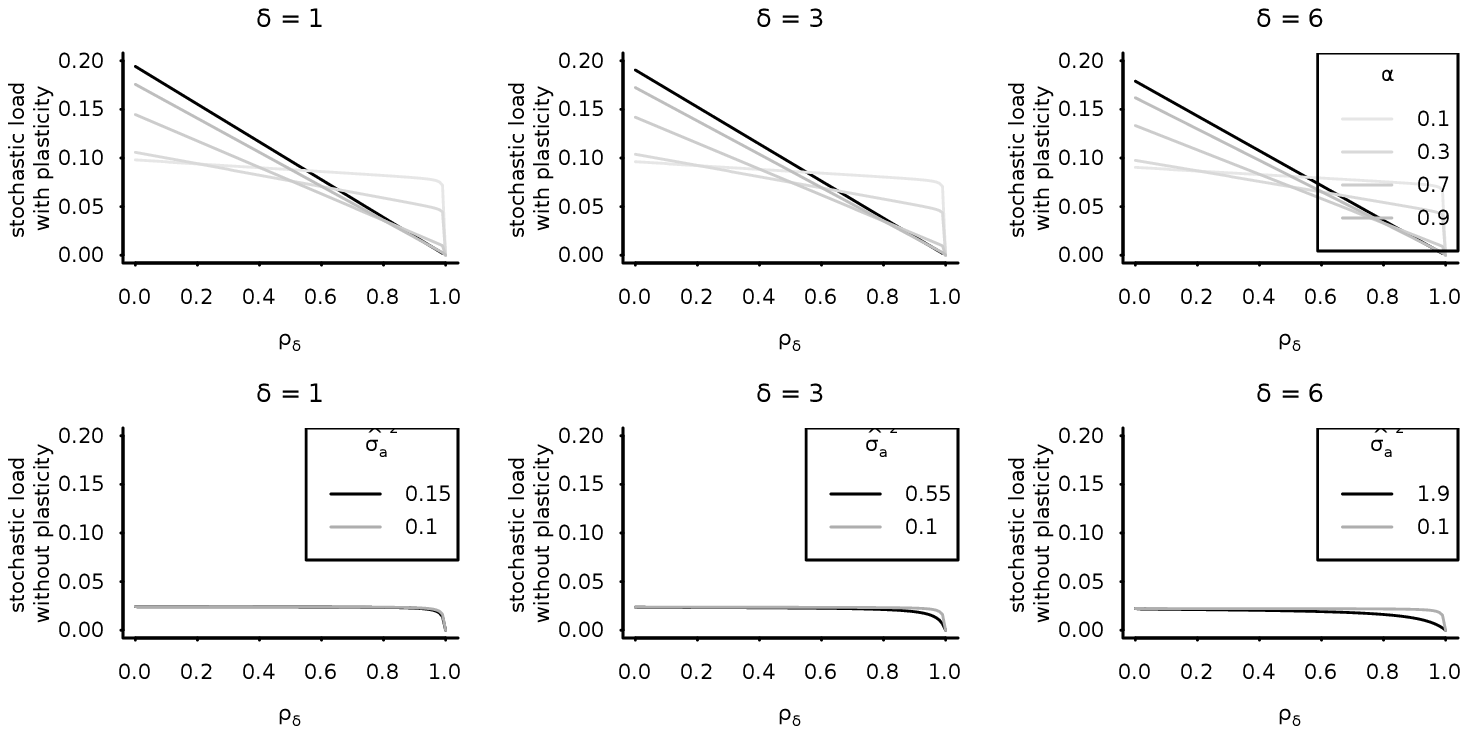}
	\caption[Stochastic load]{Top row: Stochastic load with plasticity
		plotted against $\rho_\delta$ for various values of relative plasticity $\alpha$.
		Bottom row: Stochastic load without plasticity
		(i.e., load component for ``positively auto-correlated
		fluctuations" from Table 1 of Lande and Shannon 1996) plotted against $\rho_\delta$ for
		equivalent values of total additive genetic variance.
		Specifically, the additive genetic variance in reaction norm
		elevation is adjusted to value $\hat \sigma_a^2$,  set either to
		its intial value ($\hat \sigma_a^2 = \sigma_a^2$, grey line) or
		to the total additive genetic variance in the new environment
		with plasticity in the top row ($\hat \sigma_a^2 = \delta ^2
		\sigma_b^2 + \sigma_a^2$, black line). Also, note that
		relative plasticity $\alpha$ has no effect in the bottom row).
		Other parameters are $\sigma_b^2=$0.05,
		$\tilde S(\delta) = $0.0446429, $B=$2, $\sigma^2=$
		1, $\sigma_a^2=$0.1.
	} \label{FS2stochload}
\end{center}
\end{figure}

\begin{figure}[h]
	\begin{center}
		\includegraphics{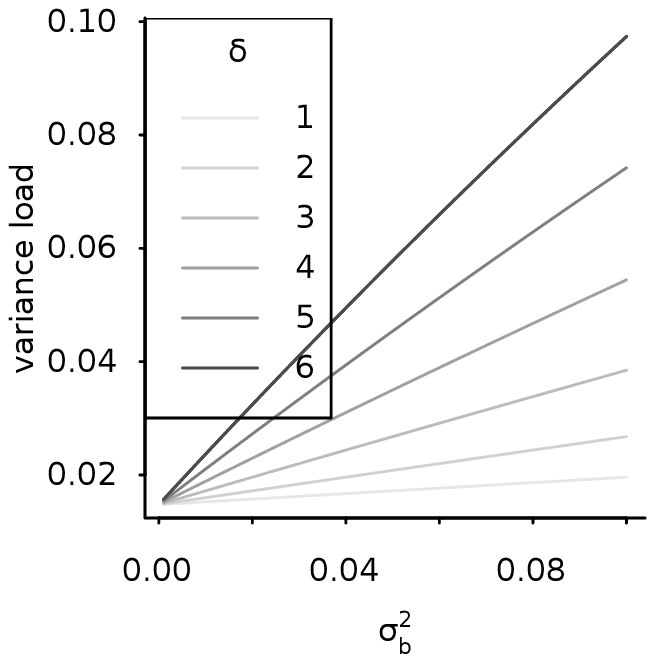}
	\caption[Variance load]{Variance load plotted against $\sigma_b^2$ for various
		values of $\delta$. Other parameters are
		$B=$2, $\sigma^2=$ 1,
		$\sigma_a^2=$0.1. Note predictability after shift
		$\rho_\delta$ does not affect variance load.
	} \label{FS3varload}
\end{center}
\end{figure}

\begin{figure}[htbp]
	\begin{center}
		\includegraphics{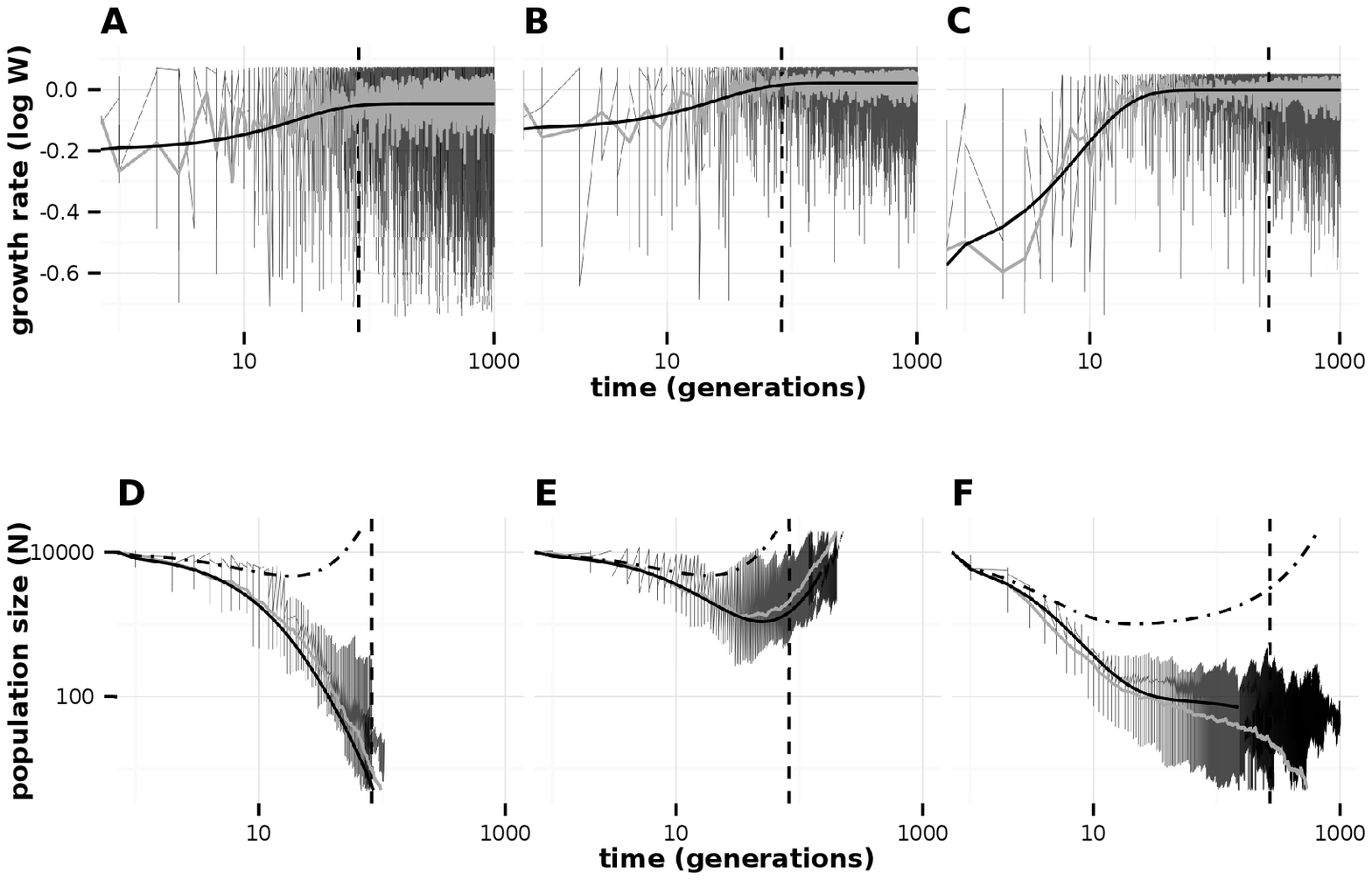}
	\caption[Dynamics]{Dynamics of growth rate ($\log \bar W$, {\bf
			A-C}), and population size ($\log N$, {\bf D-F})
		versus time (in generations, log scale) under evolutionary
		rescue for three scenarios of environmental shift $\delta$ and
		predictability $\rho_\delta$ following the shift:
		a modest shift and low predictability
		($\delta = $2.5,
		$\rho_\delta = $0.3: {\bf A,D}),
		a modest shift and high predictability
		($\delta = $2.5,
		$\rho_\delta = $0.7: {\bf B,E}), and
		a large shift and high predictability
		($\delta = $5,
		$\rho_\delta = $0.7: {\bf C,F}).
		Each panel shows 10 replicate simulations
		of 1500 generations (thin black lines), the mean of
		these simulations (thick grey line), and predicted trajectories
		of the mean (solid black line);
		also shown for comparison are predictions without amplifying
		effect of plasticity on stochastic fluctuations (dash-dot line).
		The dashed vertical lines indicate the time during Phase 1 at
		which we compute 
		quasi-extinction before rescue ($t_{\rm bef}$; see Figure~\ref{Fexplain}).
		Parameters: initial predictability $\rho =$ 0.5,
		additive genetic variance in plasticity $\sigma_b^2=$
		0.05; other parameters as in Figure 2. Greater
		shift size implies larger increase of additive genetic variance
		in the new environment; for a modest shift ({\bf A,B, D,E}) our
		model assumes additive genetic variance increases by a factor of
		4, for a large
		shift, ({\bf C,F}) the increase is by a factor of
		 14.	}
	\label{FS4dynamics}
\end{center}
\end{figure}

\begin{figure}[htbp]
	\begin{center}
		\includegraphics{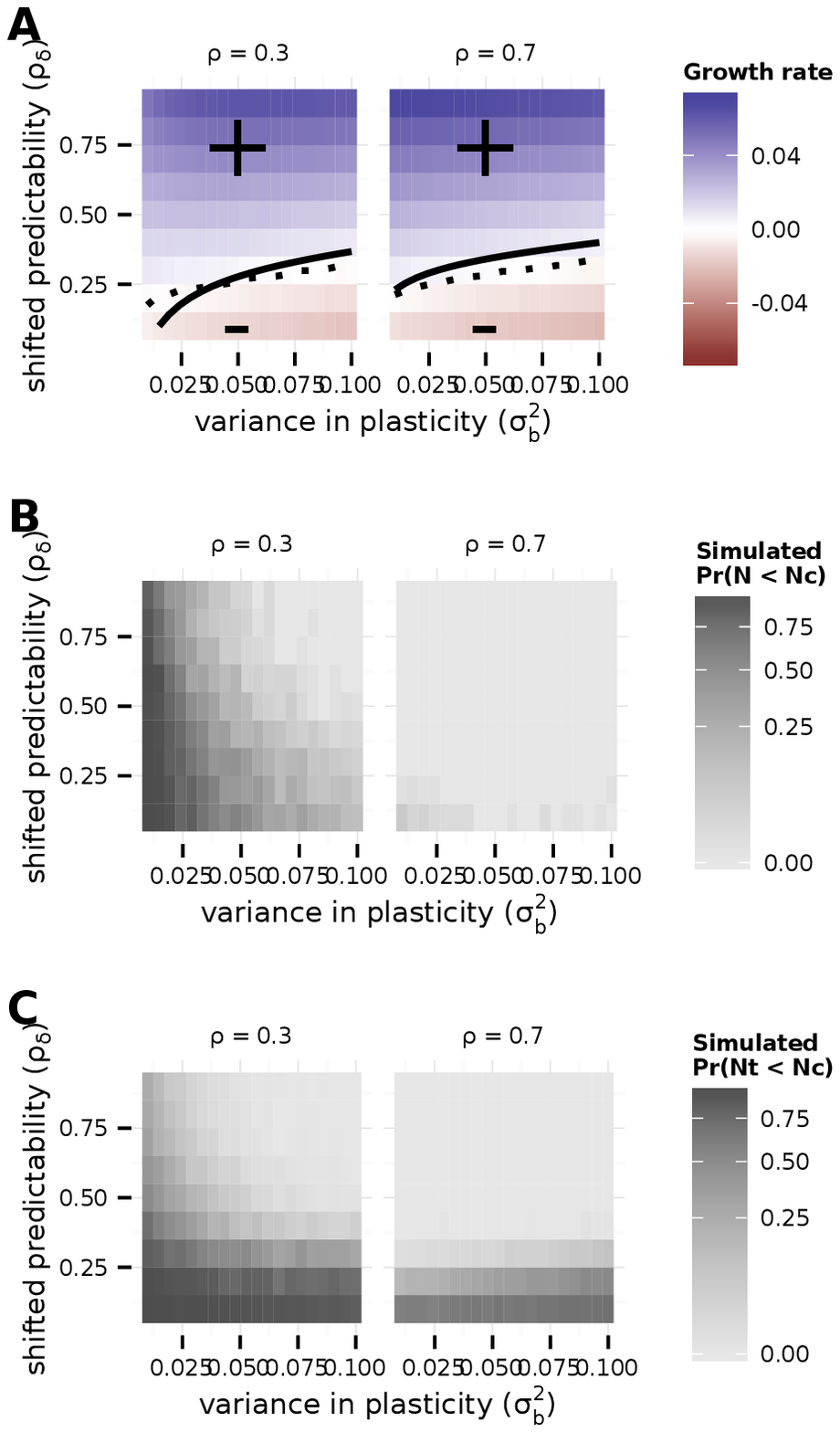}
	\caption[Predictability, additive variance in plasticity, and mean shift]{
		Potential for evolutionary rescue over a range of values for
		post-shift predictability $\rho_\delta$ versus genetic variance
		in plasticity the reference environment $\sigma_b^2$ ({\bf A,B}).
		Within panels, columns show
		low ($\rho =$ 0.3) and high  ($\rho =$
		0.7) initial predictability.  {\bf A} Growth
		rates at the end of rescue are computed from numerical
		simulations as the stochastic growth rate $\lambda_s$ between
		$t_{\rm bef}$ and $t_{\rm aft}$ spanning Phase 1 and 2
		(diverging heatmap: white 0, blue positive, and red negative).
		Black lines indicate the threshold between decline (-) and
		persistence (+) based on the analytical approximation
		($\bar r_1 = 0$, eqn~\ref{Eltpersist} ; solid line) and
		stochastic simulations ($\lambda_s = 0$, dotted black line).
		{\bf B} Simulated probability of quasi-extinction before
		rescue.  Quasi-extinction is defined at $t_{\rm bef}$ as
		illustrated in Figure~\ref{Fexplain}. 
						Shift size is set to $\delta=$ 2.5 (so that additive
		variance increases by a factor of 4 in
		the new environment when $\sigma_b^2 = $0.05).  Other parameters: initial population size
		$N(0)=$ \ensuremath{10^{4}}, selection strength $\omega^2=$
		20, developmental delay $\tau=$ 0.2,
		additive genetic $\sigma_a^2=$ 0.1 and environmental
		$\sigma_e^2=$ 0.5 variances, and maximum fitness
		$e^{r_{\max}}=$ 1.1.
	}\label{FS6potrescuesmalldelta}
\end{center}
\end{figure}

\begin{figure}[htbp]
	\begin{center}
		\includegraphics{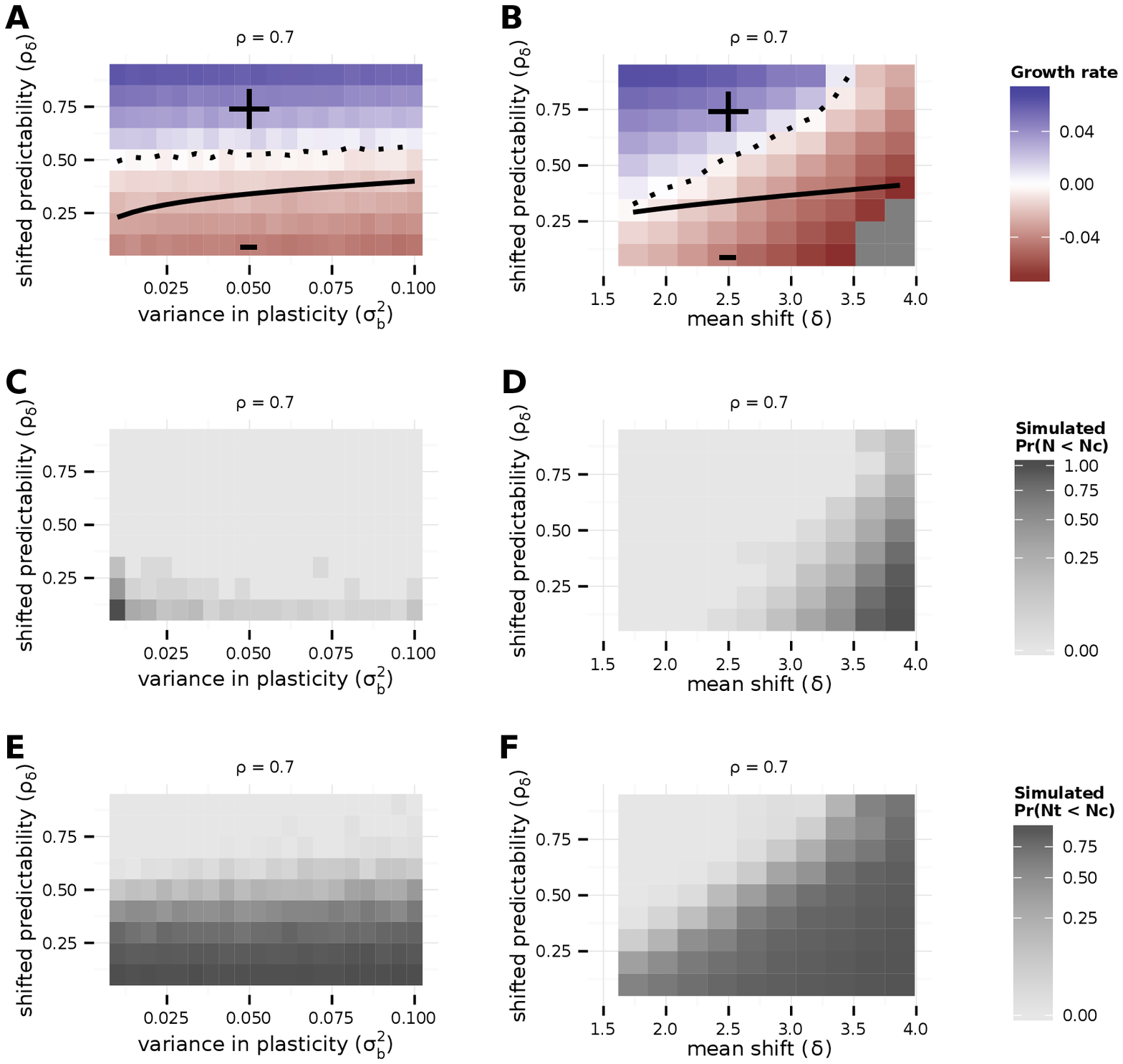}
	\caption[Predictability, varying additive variance in plasticity, and mean shift]{
		Figure \ref{Fpotrescue} but with genetic variance changing with
		population size according to a modified Stochastic House of
		Cards (SHC) approximation: $\sigma_{SHC}(\sigma_g^2) =
		\sigma_g^2 / (1 + \frac{\omega^2 + \sigma_e^2}{\mu^ 2 N_e})$ and $N_e \approx 2 R_0 N  /
		(2 R_0 - 1)$.
		Parameters and panels are as in Figure \ref{Fpotrescue} with the additional
		parameter $\mu^2$, which is the variance of the effect of
		new mutations, fixed at $\mu^2 = 0.005$.
		For the un-modified SHC the numerator, $\sigma_g^2$, is replaced by a term
		that includes the per-generation total mutation rate $V_m$ and the strength
		of selection: $2 V_m (\omega^2 + \sigma_e ^2)$ \citep[see][who
		used $\alpha^2$ instead of $\mu^2$]{Burger1995}.
		Thus, the modified version we use effectively assumes that
		larger initial values of $\sigma_a^2$ and $\sigma_b^2$ reflect
		populations with larger mutation rates.}\label{FS7potrescuevg}
\end{center}
\end{figure}

 \clearpage{}

\end{document}